\documentclass[fleqn,11pt]{article}
\usepackage{epsfig}
\input{amssym.def}
\input{amssym.tex}
\author{ }
\date{ }
\makeatletter

\def\compoundrel#1\over#2{\mathpalette\compoundreL{{#1}\over{#2}}}
\def\compoundreL#1#2{\compoundREL#1#2}
\def\compoundREL#1#2\over#3{\mathrel
      {\vcenter{\hbox{$\m@th\buildrel{#1#2}\over{#1#3}$}}}}
\begin{document}
%
\vspace*{0.5cm}
\hfill\ {\Large\bf KEK Preprint 2014-35}

\vspace*{2mm}
\hfill\ {\Large\bf November 2014\,~~~~~~~~~}

\vspace*{2mm}
\hfill\ {\Large\bf H\,~~~~~~~~~~~~~~~~~~~~~~~~~~~}

\vspace*{3.0cm}
\begin{center}
{\Huge \bf{A Multi-MW Proton/Electron Linac} }\\ 
\vspace*{0.6cm}
{\Huge \bf{at KEK} }\\
\vspace*{4.4cm}
{\LARGE R. BELUSEVIC}\\
\vspace*{0.8cm}
{\large IPNS, {\em High Energy Accelerator Research Organization} (KEK)}\\
\vspace*{1.3mm}
{\large 1-1 {\em Oho, Tsukuba, Ibaraki} 305-0801, {\em Japan}} \\
\vspace*{1.3mm}
{\large belusev@post.kek.jp}\\
\end{center}

\thispagestyle{empty}

\newpage

\tableofcontents
\addtocontents{toc}{\protect\vspace{1.3cm}}
\vspace*{5mm}
\noindent
{\large\bf References}


\newpage

\begin{center}
\begin{minipage}[t]{13.6cm}
{\bf Abstract\,:}\hspace*{3mm}
{The main `bottleneck' limiting the beam power in circular machines is caused by
space charge effects that produce beam instabilities. To increase maximally the
beam power of a `proton driver', it is proposed to build a facility consisting
solely of a 2.5\,GeV injector linac (PI) and a 20\,GeV pulsed superconducting
linac (SCL). Such a facility could be constructed using the existing KEK
accelerator infrastructure. The PI, based on the {\em European Spallation Source}
(ESS) linac, would serve both as an injector to the SCL and a source of proton
beams that could be used to copiously produce, e.g., neutrons and muons. Protons
accelerated by the SCL would be transferred through the KEK Tristan ring in order
to create neutrino, kaon and muon beams for fixed-target experiments. At a later
stage, a 70\,GeV proton synchrotron could be installed inside the Tristan ring.
The SCL, comprising 1.3\,GHz ILC-type rf cavities, could also accelerate
polarized or unpolarized electron beams. After acceleration, electrons could be
used to produce polarized positrons, or may traverse an XFEL undulator.}
\end{minipage}
\end{center}

\vspace*{0.3cm}
\renewcommand{\thesection}{\arabic{section}}
\section{Introduction}
\vspace*{0.3cm}

\setcounter{equation}{0}

~~~~The {\em Standard Model} (SM) of particle physics gives a coherent
quantum-mechanical description of electromagnetic, weak and strong interactions
based on fundamental constituents --- quarks and leptons --- interacting via
force carriers --- photons, W and Z bosons, and gluons. The SM is supported by
two theoretical `pillars': the {\em gauge principle} and the {\em Higgs
mechanism} for particle mass generation. Whereas the gauge principle has been
firmly established through precision electroweak measurements, the Higgs
mechanism is yet to be fully tested.

Preliminary results on searches for a SM Higgs boson were presented in 2012 by
the ATLAS and CMS collaborations at the CERN {\em Large Hadron Collider} (LHC)
\cite{ATLAS, CMS}. A state decaying to several distinct final states had been
observed with a statistical significance of five standard deviations. The
observed state has a mass of about 125 GeV. Its production rate is consistent
with the predicted rate for the SM Higgs boson. Event yields in different
production topologies and different decay modes are self-consistent \cite{PDG}.

To discover a new particle (such as the Higgs boson), or to search for physics
beyond the SM, usually requires the use of high-energy hadron or
electron-positron colliders. However, many important discoveries in particle
physics have been made using proton beams with relatively low energies but high
intensities (flavor mixing in quarks and in neutrinos are noteworthy examples).
Experiments with high-intensity neutrino beams, e.g., are designed primarily to
explore the mass spectrum of the neutrinos and their properties under the CP
symmetry.

Some of the most important discoveries emerged from high-precision studies of K
mesons (`kaons'), in particular {\em neutral kaons}. A deeper insight into CP
violation is expected to be gained from measurements of ultra-rare kaon decays
such as $K_{\rm L}^{0} \rightarrow \pi^{0}\nu\bar{\nu}$ and $K_{\rm L}^{+}
\rightarrow \pi^{+}\nu\bar{\nu}$. These decays provide important information on
higher-order effects in electroweak interactions, and therefore can serve as a
probe of new phenomena not predicted by the Standard Model.

The physics programs briefly described in this note are, to a large extent,
complementary to each other. For instance, neutrino oscillation experiments and
searches for permanent electric dipole moments both look for new sources of CP
violation, a phenomenon which reflects the fundamental difference between matter
and antimatter.

A unique feature of the proposed facility is the use of superconducting ILC-type
cavities to accelerate {\em both} protons and electrons, which considerably
increases its physics potential. Polarized electrons and positrons can be used
to study the structure of composite particles and the dynamics of strong
interactions, as well as to search for new physics beyond the Standard Model.

\vspace*{0.3cm}
\renewcommand{\thesection}{\arabic{section}}
\section{The Proposed Proton/Electron Facility at KEK}
\vspace*{0.3cm}

~~~~The main `bottleneck' limiting the beam power in circular machines is caused
by space charge effects that produce beam instabilities. Such a `bottleneck'
exists at the J-PARC proton synchrotron complex, and is also intrinsic to the
`proton drivers' envisaged at CERN and Fermilab. To increase maximally the beam
power of a `proton driver', it is proposed to build a facility consisting solely
of a low-energy injector linac and a high-energy {\em pulsed} superconducting
linac. Pulsed operation is preferred over the CW mode (continuous wave, 100\%
duty) mainly because the former allows the use of rf cavities with high
accelerating gradients. This would considerably reduce the overall length of the
machine, which is limited by the size of the KEK site.

\begin{figure}[h]
\vspace{3.3mm}
\begin{center}
\epsfig{file=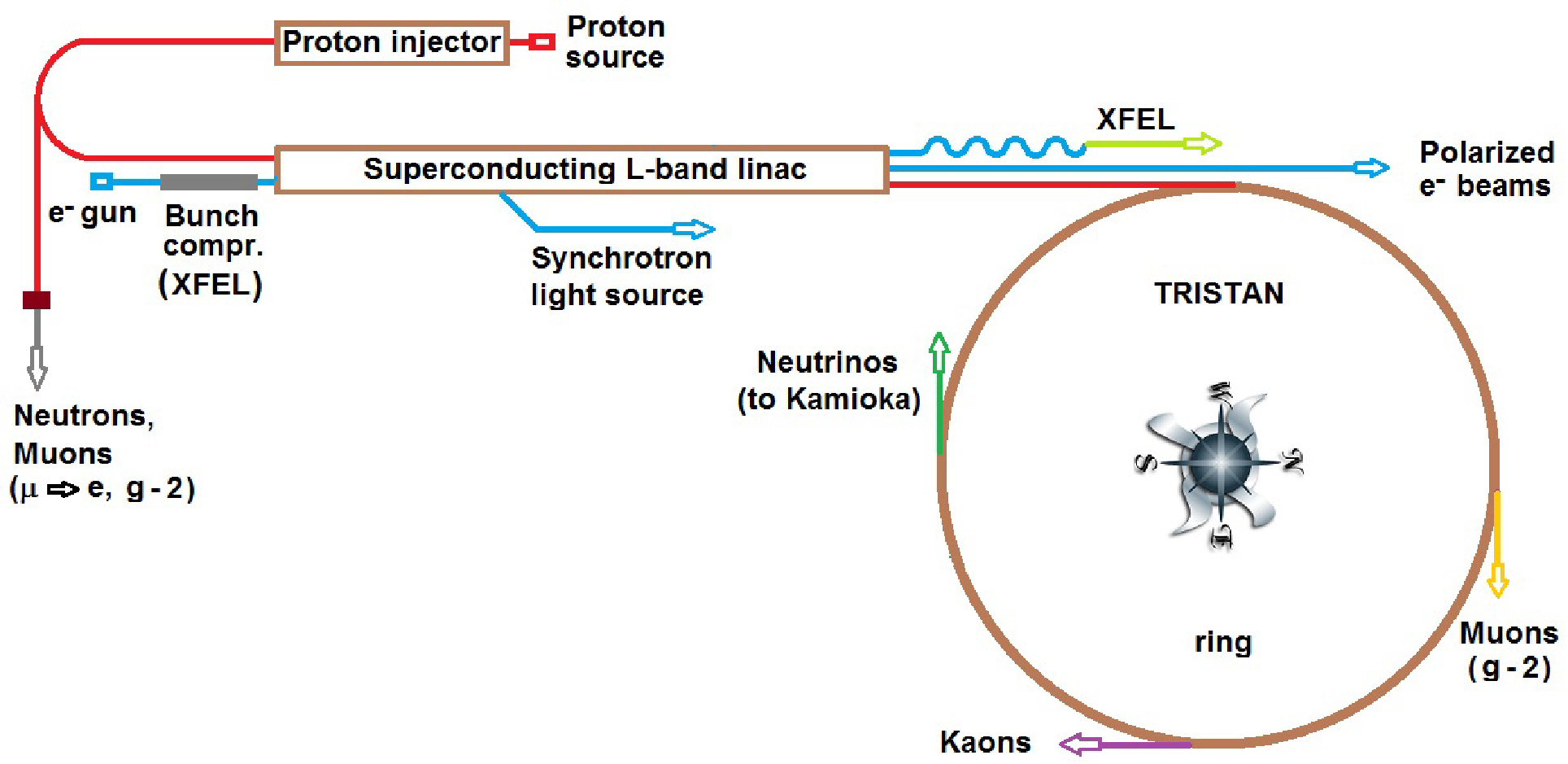,height=0.32\textheight}
\end{center}
\vskip -5mm
\caption{The layout of the proposed multi-purpose proton/electron facility at
KEK. The Tristan ring would initially be equipped only with magnets capable of
steering the proton bunches, accelerated by the superconducting linac, to
various fixed targets.}
\label{fig:Linac}
\end{figure}

The layout of the proposed proton/electron facility at KEK is shown in
Fig.\,\ref{fig:Linac}. A 2.5\,GeV proton linac (PI) serves both as an injector to
a superconducting linac (SCL) and a source of proton beams that can be
used to copiously produce neutrons and muons. Protons accelerated by the SCL to
20 GeV are transferred through the KEK Tristan ring in order to create beams for
various fixed-target experiments. At a later stage, a 70\,GeV proton
synchrotron could be installed inside the Tristan ring. The SCL, comprising
1.3\,GHz superconducting ILC-type rf cavities, can also accelerate polarized or
unpolarized electron bunches. After acceleration, electrons may traverse an
XFEL undulator, or could be used to produce polarized positrons. An SCL-based
XFEL and a synchrotron light source for applications in materials science and
medicine are also envisaged.

\begin{figure}[t]
\vspace{-3.3mm}
\begin{center}
\epsfig{file=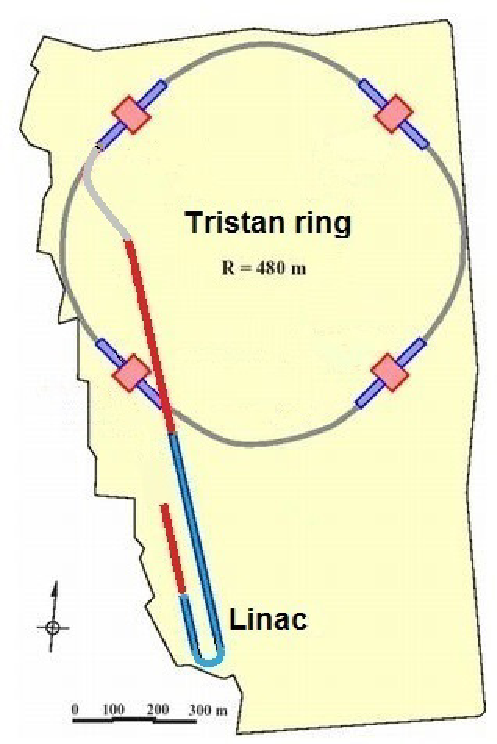,height=0.42\textheight}
\end{center}
\vskip -5mm
\caption{A sketch of the KEK site showing the Tristan ring and the existing
electron linac (in blue). The lines drawn in red indicate possible extensions
of the present linac tunnels. Alternatively, a new underground linac tunnel
could be excavated at a greater depth if demanded by radiation safety
requirements.}
\label{fig:KEK}
\end{figure}

\begin{figure}[t]
\vspace{-3.3mm}
\begin{center}
\epsfig{file=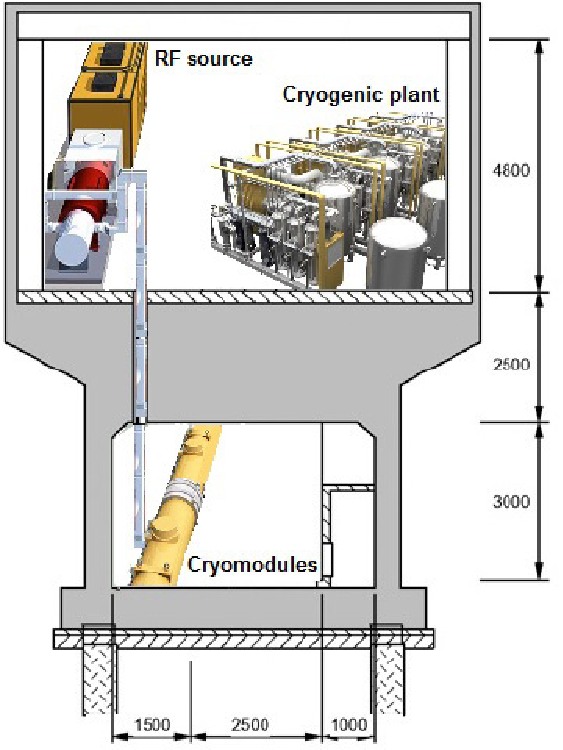,height=0.45\textheight}
\end{center}
\vskip -5mm
\caption{Front view of the linac tunnel and the klystron gallery housing the
cryomodules, RF sources and cryogenic plant units of the proposed
superconducting linac complex.}
\label{fig:Tunnel}
\end{figure}

The proposed facility would be constructed using the existing KEK accelerator
infrastructure. As shown in Fig.\,\ref{fig:KEK}, the present KEK linac tunnel
and klystron gallery could be extended to increase the length, and hence the
maximum energy, of each linac in Fig.\,\ref{fig:Linac}. The cryomodules, RF
sources and cryogenic plant units of the proposed linac complex would be
installed inside these extended structures (see Fig.\,\ref{fig:Tunnel}). The
Tristan ring (TR) would initially be equipped only with magnets capable of
steering the SCL proton bunches to various fixed targets. The four 200\,m-long
straight sections of the TR, each with an experimental hall in the middle (see
Fig.\,\ref{fig:KEK}), would house beam lines and detectors.


The beam power of a pulsed linear accelerator is given by the expression
\begin{equation}
{\cal P}_{\rm beam}^{~}\,\mbox{[MW]} = {\rm E}_{\rm b}^{~}\,\mbox{[MV]}\times
I\,\mbox{[A]}\times \tau_{\rm p}^{~}\,\mbox{[s]} \times {\cal R}\,\mbox{[Hz]}
\end{equation}
where ${\cal P}_{\rm beam}^{~}$ is the {\em beam power}, ${\rm E}_{\rm b}$ is
the {\em beam energy}, $I$ is the {\em average current per pulse}, $\tau_{\rm p}
^{~} $ is the {\em beam pulse length}, and ${\cal R}$ is the {\em repetition
rate}. The {\em duty cycle} of a pulsed linac is ${\cal D} \equiv \tau_{\rm p}
^{~}{\cal R}$. Using the values from Table 1, and assuming
${\rm E}_{\rm b} = 20$ GeV, one obtains ${\cal D} = 0.024$ and
\begin{equation}
{\cal P}_{\rm beam}^{~} = 20,000\,\mbox{MV} \times 31\,\mbox{mA} \times 1.2\,
\mbox{ms} \times 20\,\mbox{s$^{-1}$} \approx 15\,\mbox{MW}
\end{equation}

The beam parameters in Table 1 are mutually constrained by the
following relations: The number of protons per second $N = {\cal P}/{\rm E}_
{\rm b}^{~}$ and the number of protons per pulse $N_{\rm p} = N/{\cal R}$;
the average current per pulse is $I\equiv (N_{\rm p}\times 1.6\times 10^{-19}\,
{\rm C})/\tau_{\rm p}^{~}$. The klystron pulse length is the sum of the rf
cavity fill time (current dependent) and the beam pulse length: $\tau = \tau_
{\rm f}^{~} + \tau_{\rm p}^{~}$. For ILC-type cavities and $I \sim 30$ mA,
$\tau_{\rm f}^{~}\approx 0.3$ ms. Since $\tau = 1.5$ ms, the beam pulse length
$\tau_{\rm p}^{~} \approx 1.2$ ms.

\vspace*{0.3cm}
\subsection{Main Characteristics of an ILC-Type Linac}
\vspace*{0.3cm}

\vspace*{2mm}
~~~~The main characteristics of a linear accelerator are determined by the
properties of its rf source (klystrons) and accelerating cavities. For a pulsed
ILC-type superconducting linac, one of the currently available rf sources is the
{\em Toshiba} E3736 {\em Multi-Beam Klystron} \cite{yano}. This source has the
following well-tested specifications: {\em rf frequency} -- 1.3 GHz; {\em peak
rf power} -- 10 MW; {\em average power} -- 150 kW; {\em efficiency} -- 65\%;
{\em pulse length} -- 1.5 ms; {\em repetition rate} -- 10 Hz. If the repetition
rate of the Toshiba klystron is increased by a factor of two, while its peak
power is reduced by the same factor (thus keeping the average power constant)
one obtains the klystron specifications presented in Table 1. For
such a klystron, a suitable 20\,Hz {\em pulse modulator} has to be developed.

A very important parameter that determines, to a large extent, the power
conversion efficiency of a klystron is its {\em perveance}, defined by
\begin{equation}
K \equiv \frac{\raisebox{-.4ex}{$I_{0}$}}{U^{3/2}}
\end{equation}
In this expression, $I_{0}$ is the beam current and $U$ is the anode voltage.
Since there is an upper limit to the applied voltage, low perveance can be only
obtained by operating with low currents. For single-beam klystrons, this
requirement is not compatible with the need for high ouput power. With this in
mind, {\em multi-beam klystrons} (MBK) were originally developed in the 1960s
\cite{gelvich}. An MBK is a parallel assembly of low-current (low-perveance)
{\em beamlets} within a common rf structure, which efficiently generates high
output power. Using Eq. (3), the output rf power of an MBK can be expressed as
\begin{equation}
{\cal P}_{\rm k}^{~} = \eta I_{0}U = \eta KU^{5/2}
\end{equation}
where $\eta$ is the klystron {\em efficiency} and $I_{0} = N_{\rm b}I_{\rm b}$
is the total beam current; $N_{\rm b}$ is the number of beamlets and
$I_{\rm b}$ is the current carried by each beamlet. For Toshiba's E3736 MBK,
$\eta = 65$\%, $U = 116$ kV and $I_{0} = 134$ A. Hence, $K = 3.4\times 10^{-6}$
A/V$^{3/2}$, klystron's peak power ${\cal P}_{\rm k}^{~} = 10$ MW and its
average power $\overline{\cal P}_{\rm k}^{~} \equiv {\cal P}_{\rm k}^{~}\times
{\cal D} = 150$ kW. Since the klystron has six beamlets, $I_{\rm b} = 22.3$ A.

The basic properties of a 1.3GHz superconducting ILC-type cavity are presented,
e.g., in \cite{ILCcav}. There are two important parameters that characterize
rf cavities: the {\em accelerating gradient} E$_{\rm acc}$ and the {\em unloaded
quality factor} $Q_{0}^{~}$. The former is a measure of how much the energy of
a particle is increased over a given length of the linac (typically expressed
in units of MV/m), while the latter specifies how well the cavity can sustain
the stored rf power. A higher value of $Q_{0}^{~}$ implies a lower rate of power
loss realtive to the stored energy.\footnote{The Q factor of an rf cavity is
defined as $Q \equiv 2\pi\times$(energy stored/energy dissipated per cycle). For
large values of $Q$, the Q factor is approximately the number of oscillations
required for the energy of a freely oscillating system to fall off to
${\rm e}^{-2\pi}$, or 0.2\%, of its original value.} ILC-type cavities must have
a nominal $Q_{0}^{~}$ greater than $1\times 10^{10}$ (a dimensionless parameter)
at E$_{\rm acc} = 31.5$ MV/m.

\begin{figure}[t]
\vspace{-3.3mm}
\begin{center}
\epsfig{file=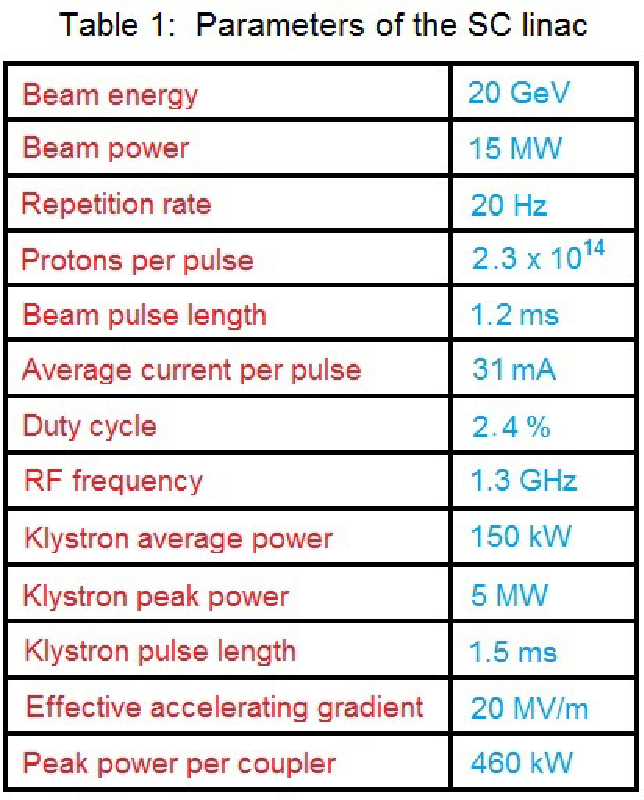,height=0.43\textheight}
\end{center}
\vskip -5mm
\label{fig:Table}
\end{figure}

Each ILC-type cryomodule for the proposed SCL would contain eight niobium 9-cell
cavities and a quadrupole magnet at its centre. Other major components of such a
cryomodule are the vacuum vessel, thermal and magnetic shields, cryogenic
piping, interconnections, etc. The inactive spaces between cavities or
cryomodules (the `packing fraction') are responsible for a substantial reduction
in the average accelerating gradient of the linac.

\begin{figure}[t]
\begin{center}
\epsfig{file=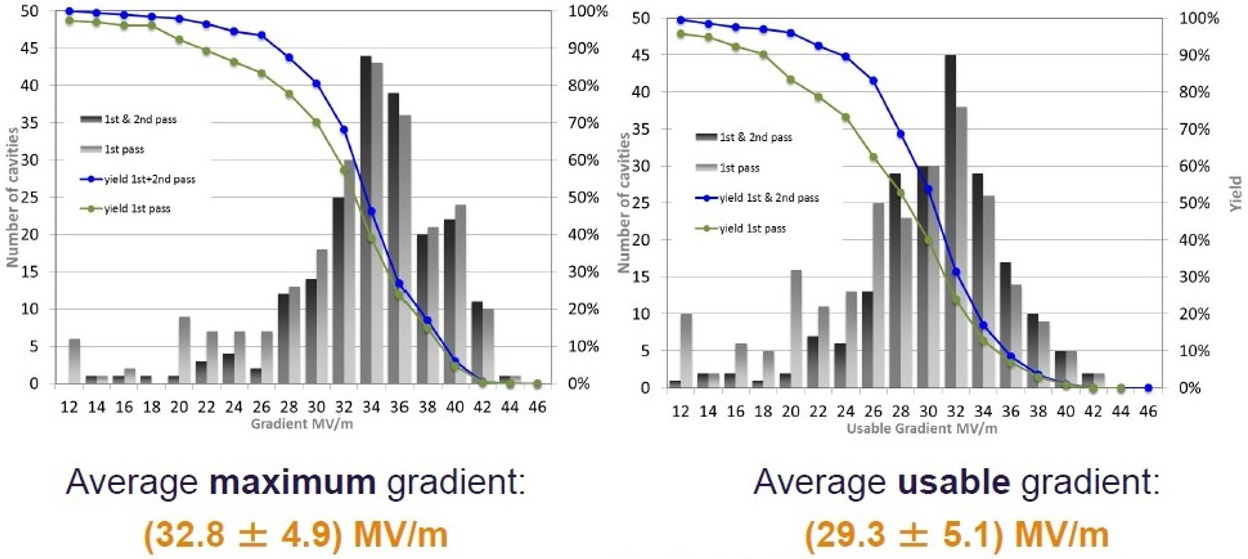,height=0.29\textheight}
\end{center}
\vskip -6mm
\caption{Measured accelerating gradients of 207 superconducting ILC cavities;
credit: D. Reschke.}
\label{fig:ILCcav}
\end{figure}

The average {\em usable} accelerating gradient in ILC-type cavities is
$\overline{\rm E}_{\rm acc} = 29.3\pm 5.1$ MV/m (see Fig.\,\ref{fig:ILCcav}).
Taking into account an estimated linac `packing fraction' of about 70\%, the
effective accelerating gradient of the SCL is E$_{\rm eff} \approx 20$ MV/m.
This implies that the total length of a 20\,GeV linac is $\sim 1000$ m.

Since the length of an ILC 9-cell cavity is 1m, a linac with ${\rm E}_{\rm b} =
20$ GeV would require $N_{\rm cav} = (20,000~{\rm MeV})/(29~{\rm MeV}) \approx
690$ cavities. Hence, the average input rf power per cavity $\overline{\cal P}_
{\rm cav}^{~} = {\cal P}_{\rm beam}^{~}/N_{\rm cav} \approx 22$ kW, and the
corresponding peak power ${\cal P}_{\rm cav}^{~} \equiv \overline{\cal P}_
{\rm cav}^{~}/{\cal D} = 916$ kW. Although this value is acceptable for a pulsed
linac with ${\cal D} \sim 2\%$, it would be prudent to use two rf couplers per
cavity. In that case the peak rf power per coupler would be about 460 kW.

For E$_{\rm acc} = 30$ MV/m, ohmic losses in an ILC 9-cell cavity amount to
${\cal P}_{\rm c}^{~} = 100$ W in the CW mode of operation, but only ${\cal P}_
{\rm c}^{~} = (100\times{\cal D})\,{\rm W} = 2.4$\,W (plus static loss) in the
pulsed mode with a duty factor ${\cal D}=0.024$. Because of large ohmic losses,
which scale with the square of the accelerating gradient, E$_{\rm acc}$ is
limited to about 15 MV/m for linacs operated in the CW mode. As already
mentioned, pulsed operation of the SCL is preferred over the CW mode mainly
because the former allows the use of rf cavities with high accelerating
gradients. This would considerably reduce the overall length of the linac, which
is limited by the size of the KEK site.

\vspace*{0.3cm}
\subsection{Proton Injector (PI)}
\vspace*{0.3cm}

\begin{figure}[t]
\vspace{-3.3mm}
\begin{center}
\epsfig{file=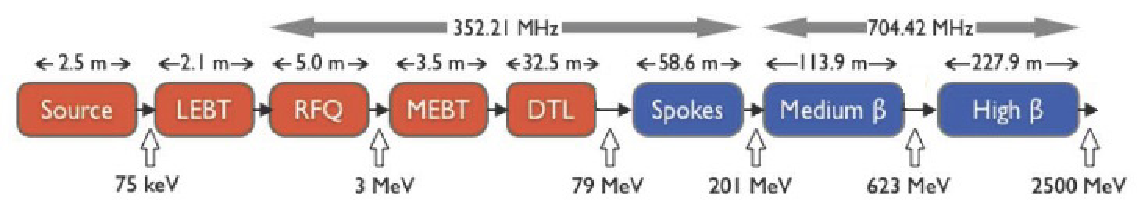,height=0.12\textheight}
\end{center}
\vskip -5mm
\caption{As an example of a typical $\sim$\,1\,GeV proton linear accelerator, a
block diagram of the ESS linac design is shown \cite{ESS}. The RFQ and DTL
structures are normal-conducting, while the spoke resonator and elliptical
cavities are superconducting. The transverse beam size along the linac varies in
the range 1--4 mm, and the bunch length decreases from 1.2\,cm to 3\,mm towards
the end of the linac.}
\label{fig:ESS}
\end{figure}

\noindent
~~~~A typical $\sim$\,1\,GeV proton linear accelerator (see Fig.\,\ref{fig:ESS})
consists of three main sections:

\vspace*{2mm}
\noindent
$\bullet$~~{\large{\em Front end}}, comprising a proton source and a
radiofrequency quadrupole accelerator;

\vspace*{1mm}
\noindent
$\bullet$~~{\large{\em Medium-velocity linac}}, which accelerates proton
beams to $\sim$\,100 MeV;

\vspace*{1mm}
\noindent
$\bullet$~~{\large{\em High-velocity linac}}, which accelerates protons to
energies exceeding 1 GeV.
\vspace*{2mm}

The most complex part of a proton linac is the low-energy (low-$\beta$) section,
situated between the proton (or ion) source and the first drift-tube-based
accelerating section. The continuous beam of protons coming from an
{\em electron cyclotron resonance} (ECR) source \cite{faircloth} has to be
focused, bunched and accelerated in the first rf structure. These three
essential functions are nowadays successfully performed by {\em radio frequency
quadrupoles} (RFQ) \cite{kapchinsky}. However, the beam has to be shrunk before
it can be fed into an RFQ. This is accomplished within a low-energy beam
transport (LEBT) section by means of cylindrical magnets (solenoids).

As soon as the beam is bunched --- which is essential for further acceleration
--- it enters a medium-energy beam transport (MEBT) section, where it is
collimated and steered from the RFQ into the medium-velocity linac (MVL). The
MEBT may also contain a number of buncher cavities. Inside the MVL, the beam is accelerated to about 100 MeV ($\beta \sim 0.1$ to 0.5). The MVL usually contains
normal-conducting {\em drift-tube linac} (DTL) and {\em cell-coupled drift tube
linac} (CCDTL) structures. A DTL incorporates accelerating components of
increasing length in order to match precisely the increase in beam velocity,
while quadrupole magnets provide strong focusing. The main advantage of using
CCDTL structures is that they provide longitudinal field stability.

High-velocity linac (HVL) structures accelerate the beam to energies around
1 GeV. They consist either of normal-conducting {\em side-coupled linac} (SCL)
structures\footnote{The main reason for using these $\pi$/2-mode structures is
that long chains of coupled cavities are often required for an efficient use
of high-power rf sources \cite{vretenar}.} or superconducting {\em elliptical
cavities}. The latter offer some advantages over the former, such as higher
accelerating gradients and lower operating costs. The superconducting HVL can
also feature {\em spoke resonators}, characterized by their simplicity, high
mechanical stability and compact size \cite{vretenar}.

One of the main concerns in the design of a high-power proton linac is to
restrict beam losses. A careful beam dynamics study is therefore needed in order
to avoid halo formation, a major source of beam loss. Another important issue is
the preservation of beam emittance \cite{mosnier}.

High-power proton linear accelerators have a wide range of applications
including spallation neutron sources, nuclear waste transmutation, production of
radioisotopes for medical use, etc. A number of laboratories worldwide have
expressed interest in building `proton drivers' that would primarily deliver
high-intensity neutrino, kaon and muon beams \cite{garoby, baussan}. The physics
potential of such a facility is discussed in the next section.


\vspace*{0.3cm}
\renewcommand{\thesection}{\arabic{section}}
\section{Physics at the Proposed Facility}
\vspace*{0.3cm}

~~~~The physics potential of a multi-MW `proton driver' is extensively
discussed, for instance, in \cite{ProjectX}. This note is mainly concerned with
the application of a high-intensity proton source to investigate the properties
of long-baseline neutrino oscillations. A unique feature of the proposed
facility is the use of a superconducting linac to accelerate {\em both} protons
and electrons, which considerably increases its physics potential. As described
in \cite{JLab, voutier}, polarized electron and positron beams can be used to
study the structure of composite particles and the dynamics of strong
interactions, as well as to search for new physics beyond the Standard Model.

\vspace*{0.3cm}
\subsection{Neutrino Flavor Oscillations and Leptonic CP Violation}
\vspace*{0.3cm}

~~~~The universe contains about a billion neutrinos for every quark or electron.
The three known neutrino species (`flavors') are named after their partner leptons in the Standard Model: {\em electron neutrino}, {\em muon neutrino} and
{\em tau neutrino}. The observed transformation of one neutrino flavor into
another (`{\em neutrino oscillation}') indicates that these ubiquitous particles
have finite masses \cite{PDG}. Cosmological data suggest that the combined mass
of all three neutrino species is a million times smaller than that of the
next-lightest particle, the electron \cite{belusevic}.

The phenomenon of neutrino oscillations implies not only the existence of
neutrino mass, but also of {\em neutrino mixing}. That is, the neutrinos of
definite flavor are not particles of definite mass (mass eigenstates), but
coherent quantum-mechanical superpositions of such states. Converesely, each
neutrino of definite mass is a superposition of neutrinos of definite flavor.
Neutrino mixing is large, in striking contrast to quark mixing. {\em Whatever
the origin of the observed neutrino masses and mixings, it implies a profound
modification of the Standard Model.}

\begin{figure}[t]
\vspace{-3.3mm}
\begin{center}
\epsfig{file=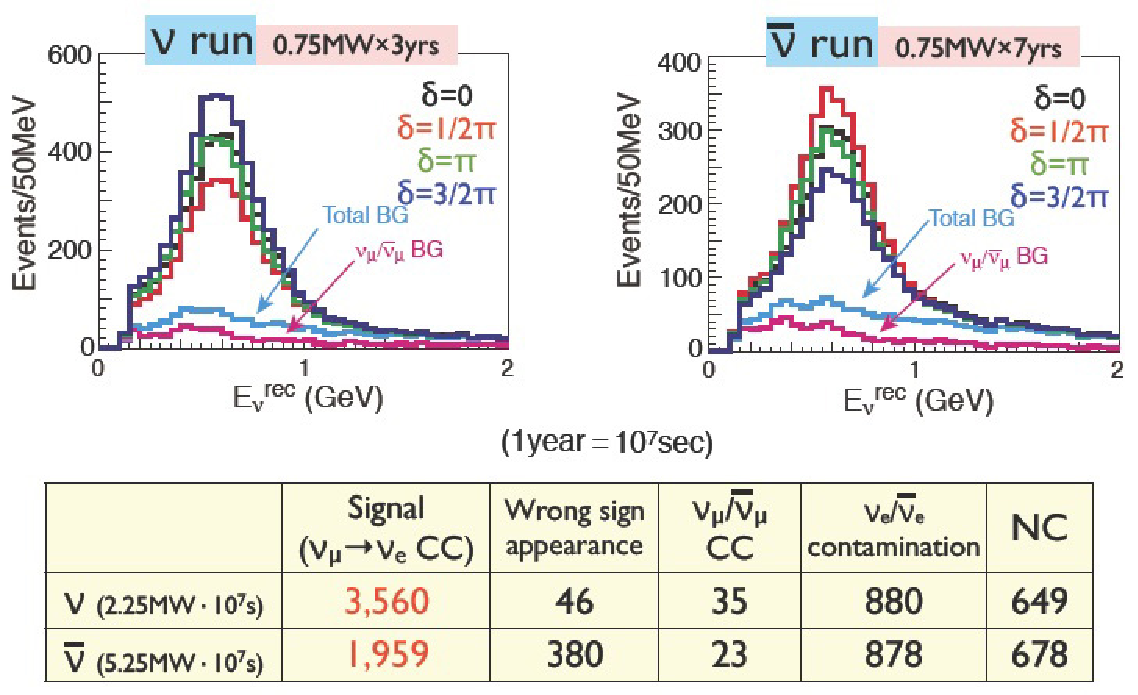,height=0.36\textheight}
\end{center}
\vskip -5mm
\caption{Predicted yield of $\nu_{e}^{~}$ and $\bar{\nu}_{e}^{~}$ `appearance
events' inside the 0.5\,Mt fiducial volume of the proposed Hyper-Kamiokande (HK)
detector, assuming that the beam power of the 30\,GeV proton synchrotron at
J-PARC is increased to 0.75 MW ($\delta$ is a CP-violating parameter); credit:
the T2K Collaboration. Over the same data taking period, beams produced at the
proposed 15\,MW `proton driver' would yield roughly the same number of
`appearance events' inside the 0.022\,Mt fiducial volume of the existing
Super-Kamiokande detector (the same baseline as for HK).}
\label{fig:Sakashita}
\end{figure}

Mathematically, the phenomenon of neutrino mixing can be expressed as a unitary
transformation relating the flavor and mass eigenstates. The neutrino
oscillation rate depends, in part, on (1) the difference between neutrino masses
and (2) the three parameters in the transformation matrix known as {\em mixing
angles}. The complex phase factors in the transformation matrix (also called
{\em mixing matrix}) are associated with the violation of CP symmetry in the
lepton sector. The size of the CP violation is determined both by the phases and
the mixing angles.

Experiments with high-intensity neutrino beams are designed primarily to
explore the mass spectrum of the neutrinos and their properties under the CP
symmetry, and thus provide a deeper insight into the nature of these elusive
particles and their role in the universe. For instance, if there is experimental
evidence for CP violation in neutrino oscillations, it could be used to explain
the observed asymmetry between matter and antimatter \cite{shaposhnik}.

The proposed T2HK (Tokai-to-Hyper-Kamiokande) project is a natural extension of
the T2K (Tokai-to-Super-Kamiokande) neutrino oscillation experiment \cite{abe}. Hyper-Kamiokande (HK), a water Cherenkov detector with a fiducial mass of about
0.5 million metric tons (0.5\,Mt), would serve as a far detector for neutrino
beams produced at the J-PARC accelerator complex, situated at a distance L$_{1}
\approx 300$ km from Kamioka. Since the baseline L$_{1}$ is relatively short,
the matter-induced neutrino mixing is rather small, which implies a weak
sensitivity to the neutrino mass hierarchy. The T2HK Collaboration intends to
resolve this hierarchy by virtue of matter-enhanced oscillations of atmospheric
neutrinos traversing the Earth \cite{abe, cahn}.

Under the assumption that the beam power of the 30\,GeV proton synchrotron at
J-PARC can be increased to 0.75 MW, the T2HK experiment would deliver about 5500
$\nu_{e}^{~}$ and $\bar{\nu}_{e}^{~}$ `appearance events' in the HK detector
within ten years of data taking (see Fig.\,\ref{fig:Sakashita}). {\em The same
event yield would be obtained within a year using the} HK {\em detector and the
proposed} KEK `{\em proton driver}'.

Alternatively, a 100\,kiloton water Cherenkov detector could be built at
Okinoshima, located along the T2K beamline at a distance L$_{2} \approx 650$ km
from KEK. Using the proposed `proton driver', the detector at Okinoshima as well
as Super-Kamiokande, the neutrino mass hierarchy could be determined either by
comparing the $\nu_{e}^{~}$ appearance probabilities measured at the two vastly
different baseline lengths L$_{1}$ and L$_{2}$, or by measuring at L$_{1}$ and
L$_{2}$ the neutrino energy of the first oscillation maximum. Once the mass
hierarchy is determined, the CP-violating phase in the mixing matrix can be
measured with a precision of $\pm 20^{\circ}$, assuming that $2.5\times 10^{21}$
protons are delivered on target for both $\nu_{e}$ and $\bar{\nu}_{e}$ beams
\cite{hagiwara}.

The main challenge in the design of a multi-MW neutrino beam facility is to
build a {\em proton target} that could dissipate large amounts of deposited
energy, withstand the strong pressure waves created by short beam pulses, and
survive long-term effects of radiation damage. Simulation studies of the pion
production and energy deposition in different targets (liquid mercury jet,
tungsten powder jet, solid tungsten bars and gallium liquid jet) are presented
in \cite{back}. Those studies also provided estimates of the amount of concrete
shielding needed to protect the environment from the high radiation generated by
each target. A proof-of-principle demonstration of a 4\,MW target station comprising a liquid mercury jet inside a 20\,T solenoidal magnetic field is
described in \cite{mcdonald}. A 15\,MW proton beam could be separated by a
series of magnets into four beam lines. Each of the four beams would be focused
by a series of quadrupoles and correctors to an assembly consisting of four
targets and the same number of magnetic horns (see, e.g., \cite{baussan1}).

To maximize the discovery potential of a neutrino beam facility, it is important
to properly design the {\em magnetic horn} that focuses the charged particles
produced in the proton target. For proton beam pulses lasting 1 ms, a DC horn
has been designed by Yukihide Kamiya of KEK \cite{kamiya}. The toroidal magnetic
field of the horn, characterized by B$(r) =$ const., is generated by hollow
aluminium conductors containing water. The strength of the magnetic field
B~$= 0.2$ T, and its length $\ell = 5$ m; hence, B$\,\cdot\,\ell =$ 1 T$\,
\cdot\,$m. The radius of the magnet, $r$, is determined by $r = L\tan (\theta )
+ \ell\tan (\theta /2)$, where $\theta\approx 0.03 + 0.3/p$ is the initial angle
a charged pion makes with respect to the proton beam direction, $L$ is the
distance from the target to the horn, and $p$ is the pion momentum. For example,
if $L = 5$ m then $r \approx 5$ m. The total power generated in the conductors is
about 10 MW.

\vspace*{0.3cm}
\subsection{Physics with Polarized Electrons and Positrons}
\vspace*{0.3cm}

~~~~Electron and positron beams, polarized and/or unpolarized, can be used to
study the structure of composite particles and the dynamics of strong
interactions, as well as to search for new physics beyond the Standard Model.
A detailed description of the physics potential of a facility that can provide
such beams (e.g., the upgraded CEBAF facility at Jefferson Lab or the proposed
KEK superconducting linac) is presented in \cite{JLab, voutier}.

{\em Polarized positrons} are created in a conversion target by circularly
polarized photons, which themselves are produced when polarized laser light is
Compton-backscattered on a high-energy electron beam \cite{omori}. Circularly
polarized photons can also be produced by bremsstrahlung from polarized
electrons \cite{grames}.  Using polarized electrons and positrons, the
{\em nucleon electromagnetic form factors} and {\em generalized parton
distributions} can be determined in a model-independent way \cite{voutier}.

Among the physics topics discussed in \cite{JLab}, parity violation in
electron-electron (M\o ller) scattering is of particular interest. M\o ller
scattering is a purely leptonic process that allows high-precision tests of the
Standard Model. At four-momentum transfers much smaller than the mass of the Z
boson ($q^{2} \ll {\rm M}_{\rm Z}^{2}$), the parity-violating asymmetry,
${\cal A}$, is dominated by the interference between the electromagnetic and
neutral weak amplitudes \cite{zeldovich}. By definition,
\begin{equation}
{\cal A} \equiv \frac{\raisebox{-.4ex}{${\rm d}\sigma_{\mbox{\tiny{\rm R}}} -
{\rm d}\sigma_{\mbox{\tiny{\rm L}}}$}}{{\rm d}\sigma_{\mbox{\tiny{\rm R}}} +
{\rm d}\sigma_{\mbox{\tiny{\rm L}}}} \approx \frac{\raisebox{-.4ex}{$f_{\mbox
{\tiny{\rm Z}}}^{\mbox{\tiny{\rm R}}} - f_{\mbox{\tiny{\rm Z}}}^{\mbox{\tiny
{\rm L}}}$}}{f_{\gamma}^{~}}
\end{equation}
In this expression, ${\rm d}\sigma_{\mbox{\tiny{\rm R}}}$ (${\rm d}\sigma_
{\mbox{\tiny{\rm L}}}$) is the differential cross-section for right-handed
(left-handed) electron scattering on an unpolarized target:
\begin{equation}
{\rm d}\sigma_{\mbox{\tiny{\rm R,L}}} \propto |f_{\gamma}^{~} + f_{\mbox{\tiny
{\rm Z}}}^{\mbox{\tiny{\rm R,L}}}|^{2} \approx |f_{\gamma}^{~}|^{2} +
2f_{\gamma}^{~}f_{\mbox{\tiny{\rm Z}}}^{\mbox{\tiny{\rm R,L}}}
\end{equation}
where $f_{\gamma}^{~}$ and $f_{\mbox{\tiny{\rm Z}}}^{\mbox{\tiny{\rm R,L}}}$ are
the scattering amplitudes with $\gamma$ and $Z$ exchange, respectively. From the
four Feynman diagrams in Fig.\,6 of \cite{marciano}, one can readily obtain the
Born amplitudes for M\o ller scattering mediated by photons and Z bosons. The
weak neutral current amplitudes are functions of the weak mixing (or Wienberg)
angle $\theta_{\rm w}^{~}$, which relates the weak coupling constants $g_{\rm w}
^{~}$ and $g_{\mbox{\tiny{\rm Z}}}^{~}$ to the electromagnetic coupling
constant. As shown in \cite{marciano}, the polarization asymmetry for polarized
electron scattering on an unpolarized target is given by
\begin{equation}
{\cal A}^{\rm Born} = m_{e}^{~}{\rm E}\,\frac{\raisebox{-.4ex}{$G_{\rm F}^{~}
Q^{e}_{\rm w}$}}{\sqrt{2}\pi\alpha}{\cal F}(\theta )
\end{equation}
where $m_{e}^{~}$ is the mass of the electron, E is the incident beam energy,
$G_{\rm F}^{~}$ is the Fermi coupling constant characterizing the strength of
the weak interaction, $\alpha$ is the fine structure constant, and ${\cal F}
(\theta )$ is a function of the scattering angle in the center-of-mass frame.
The {\em weak charge} of the electron, $Q^{e}_{\rm w} = 1 - 4\sin^{2}\theta_
{\rm w}^{~}$, is proportional to the product of the electron's vector and
axial-vector couplings to the Z boson.

\begin{figure}[t]
\vspace{-3.3mm}
\begin{center}
\epsfig{file=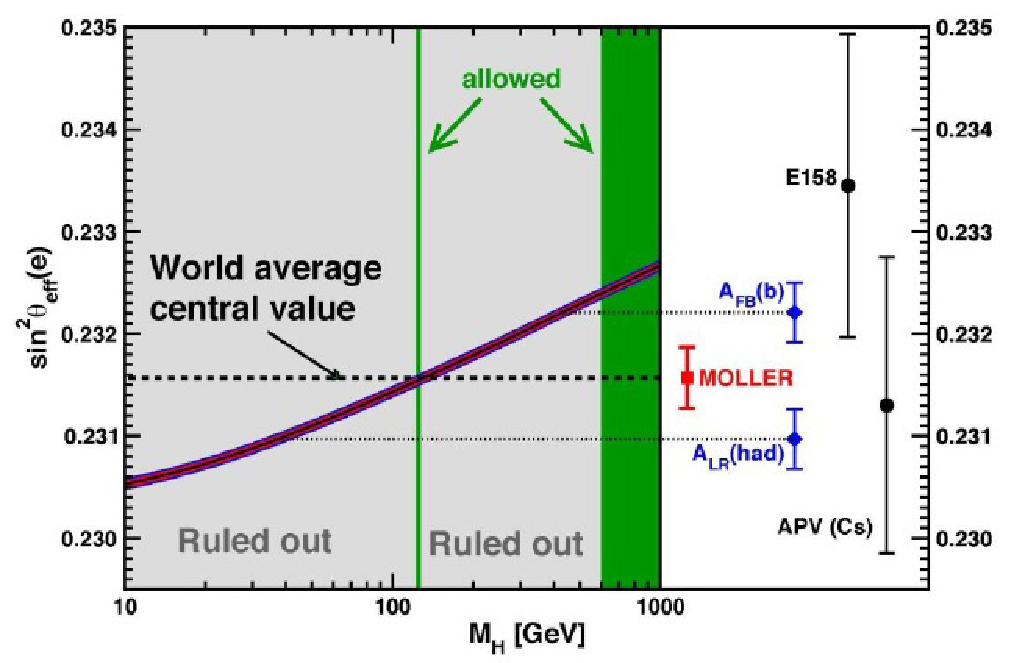,height=0.31\textheight}
\end{center}
\vskip -6mm
\caption{Indirect constraints on the Higgs-boson mass, M$_{\rm H}^{~}$, from the
most precise high $q^{2}$ (blue) and low $q^{2}$ (black) measurements of
$\sin^{2}\theta_{\rm w}^{~}$, evolved to the same energy scale ($q^{2} = {\rm M}
_{\rm Z}^{2}$); credit: Jens Erler and \cite{JLab}. {\em Green bands}: allowed
M$_{\rm H}^{~}$ regions; {\em Red curve}: theory predictions for $\sin^{2}
\theta_{\rm w}^{~}$ vs M$_{\rm H}^{~}$; {\em Dashed line}: average value of
$\sin^{2}\theta_{\rm w}^{~}$ from all electroweak asymmetry data; {\em Red}:
expected measurement precision (placed at an arbitrary $y$-axis value) of the
MOLLER experiment \cite{moller}, in which a 11\,GeV longitudinally polarized
electron beam would scatter on atomic electrons in a liquid hydrogen target.}
\label{fig:JLab}
\end{figure}

Since the value of $\sin^{2}\theta_{\rm w}^{~}$ is close to 1/4, there is an
enhanced sensitivity of ${\cal A}$ to small changes in the Weinberg angle. The
value of $\theta_{\rm w}^{~}$ varies as a function of the four-momentum
transfer, $q$, at which it is measured. This variation, or `running', is a key
prediction of the Standard Model. The one-loop {\em electroweak radiative
corrections} to ${\cal A}^{\rm Born}$ (calculated once the renormalized
parameters in (7) are properly defined) reduce its Born value by $\sim$40\%.
This effect can be attributed to an increase of $\sin^{2}\theta_{\rm w}^{~}
(q^{2})$ by 3\% as the four-momentum transfer `runs' from $q^{2} = {\rm M}_
{\rm Z}^{2}$ to $q^{2} \approx 0$ \cite{marciano1}.

Many of the electroweak meaurements obtained over the past three decades may be
combined to provide a global test of consistency with the SM. Since the
Higgs-boson mass affects the values of electroweak observables through radiative
corrections, it is of fundamental importance to test the agreement between the
directly measured value of M$_{\rm H}^{~}$ and that inferred from the
measurements of electroweak parameters $\sin^{2}\theta_{\rm w}^{~}$ (see
Fig.\,\ref{fig:JLab}), M$_{\rm W}^{~}$ and M$_{\rm top}^{~}$. High-precision
electroweak measurements, therefore, represent a natural complement to direct
studies of the Higgs sector.

Apart from providing a comprehensive test of the SM, precision measurements of
weak neutral current interactions at $q^{2} \ll {\rm M}_{\rm Z}^{2}$ also allow
indirect access to new physics phenomena beyond the TeV energy scale. For
instance, such measurements can be used to look for hypothetical Z' bosons,
4-fermion contact interactions, or very weekly coupled low-mass `dark bosons'
\cite{kumar}.

\subsection{Rare Kaon Decays}
\vspace*{0.3cm}

~~~~Some of the most important discoveries in particle physics emerged from
studies of K mesons (`kaons'), in particular {\em neutral kaons}. The neutral K
meson, $K^{0}$, and its antiparticle, $\bar{K}^{0}$, form a remarkable
quantum-mechanical two-state system that has played a significant role in the
development of the Standard Model \cite{belusev}.

In 1953, it was shown that the observed $2\pi$ and $3\pi$ decay modes of the
charged kaon required the parent particles to have opposite intrinsic parities.
This suggested that parity may not be conserved. Three years later, experiments
demonstrated that parity was indeed violated in weak interactions. The first
indication that parity violation was accompanied by a failure of charge
conjugation was seen in 1964, when one $2\pi$ event was detected for every 500
or so common $3\pi$ decays of the long-lived neutral kaon, $K_{\rm L}^{0}$. The
concept of {\em strangeness}, introduced in 1953 to explain the anomalously long
lifetimes of K mesons, was crucial for the development of the quark model of
particles. In 1970, the smallness of the observed branching ratio for
$K_{\rm L}^{0} \rightarrow \mu^{+}\mu^{-}$, implying the absence of
strangeness-changing neutral currents, led to the prediction of a fourth quark,
the {\em charm} quark. The sensitivity of $K^{0}$-$\bar{K}^{0}$ mixing to
energies higher than the kaon mass scale was used soon thereafter to predict the
mass of the charm quark (discovered in 1974).

CP violation was introduced in the SM by increasing the number of quark and
lepton families to at least three (M. Kobayashi and T. Maskawa, 1973). This idea
became very attractive with the subsequent discovery (in 1977) of the
{\em bottom} quark, which forms, together with the {\em top} quark (discovered
in 1995), a third family of quarks. It is a remarkable property of the
Kobayashi-Maskawa model that {\em quark mixing} and CP violation are intimately
related.

\begin{figure}[h]
\begin{center}
\epsfig{file=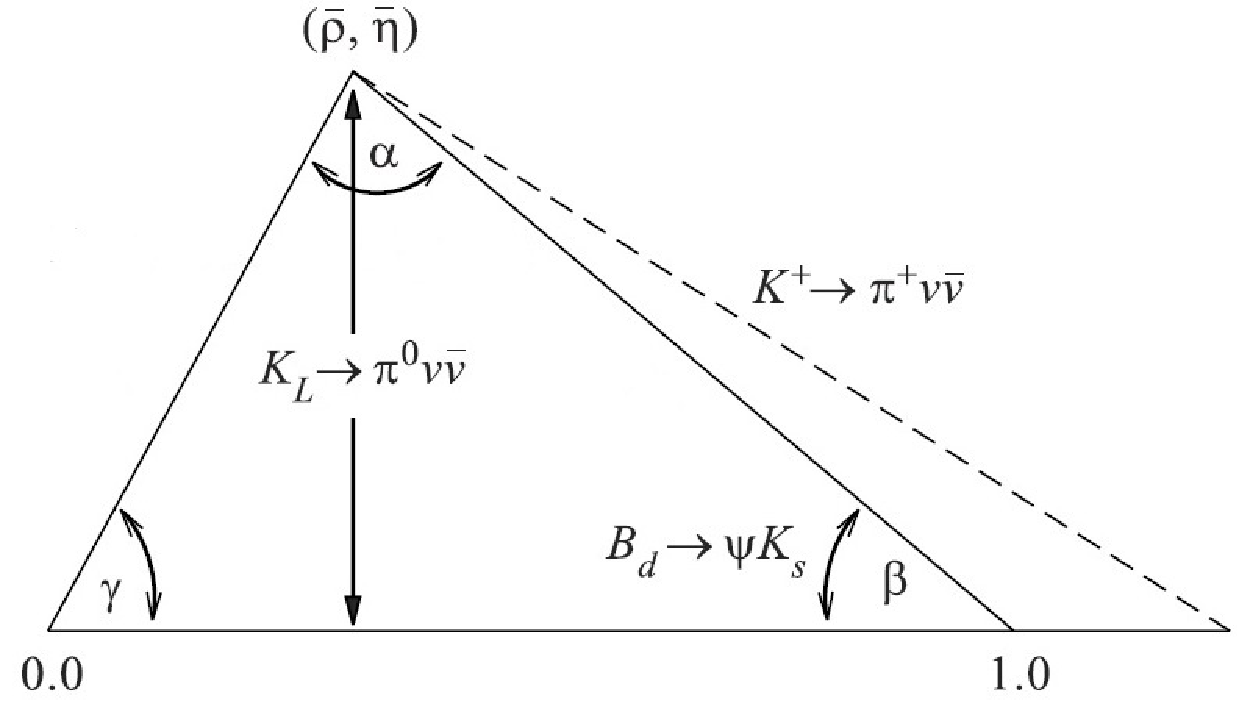,height=0.25\textheight}
\end{center}
\vskip -6mm
\caption{Unitarity triangle from $K \rightarrow \pi\nu\bar{\nu}$ decays. The
displacement of the bottom-right vertex is due to the charm-quark contribution
to $K^{+} \rightarrow \pi^{+}\nu\bar{\nu}$.}
\label{fig:Unitarity}
\end{figure}

A deeper insight into CP violation is expected to be gained from precision
measurements of rare kaon decays such as $K_{\rm L}^{0} \rightarrow\pi^{0}\nu
\bar{\nu}$ and $K^{+} \rightarrow \pi^{+}\nu\bar{\nu}$. Both decays are
theoretically `clean' because hadronic transition amplitudes are matrix elements
of quark currents between mesonic states, which can be extracted from the
leading semileptonic decays using isospin symmetry. Since photons do not couple
to neutrinos, $K \rightarrow \pi\nu\bar{\nu}$ decays are entirely due to
second-order weak processes determined by Z-penguin and W-box diagrams
\cite{belusev}.

The process $K_{\rm L}^{0} \rightarrow \pi^{0}\nu\bar{\nu}$ proceeds almost
entirely through direct CP violation, and is completely determined by
`short-distance' one-loop diagrams with top quark exchange. The Standard Model
predicts its branching ratio to be ${\cal B}(K_{\rm L}^{0} \rightarrow \pi^{0}
\nu\bar{\nu}) = (2.43\pm 0.39)\times 10^{-11}$ \cite{brod}. This decay is an
important source of information on higher-order effects in electroweak
interactions, and therefore can serve as a probe of physics beyond the Standard
Model (see \cite{belyaev} and references therein).

\newpage

The decay $K^{+} \rightarrow \pi^{+}\nu\bar{\nu}$ receives both CP-conserving
and CP-violating contributions. It has theoretical uncertainties that are
somewhat larger than those in the process $K_{\rm L}^{0}\rightarrow \pi^{0}
\nu\bar{\nu}$. Since both decays involve one-loop Feynman diagrams with top
quark exchange, they can yield valuable measurements of the CKM matrix elements
$|V_{td}^{~}|$ and $|V_{ts}^{~}|$. The quantity Im$V_{ts}^{*}V_{td}^{~}$, which
can be obtained from $K_{\rm L}^{0} \rightarrow \pi^{0}\nu\bar{\nu}$ alone,
plays a central role in the phenomenology of CP violation in K decays; this
quantity is related to the {\em Jarlskog parameter}, the invariant measure of
CP violation in the Standard model \cite{belusev, buras}.
 
\begin{figure}[t]
\vspace{-3.3mm}
\begin{center}
\epsfig{file=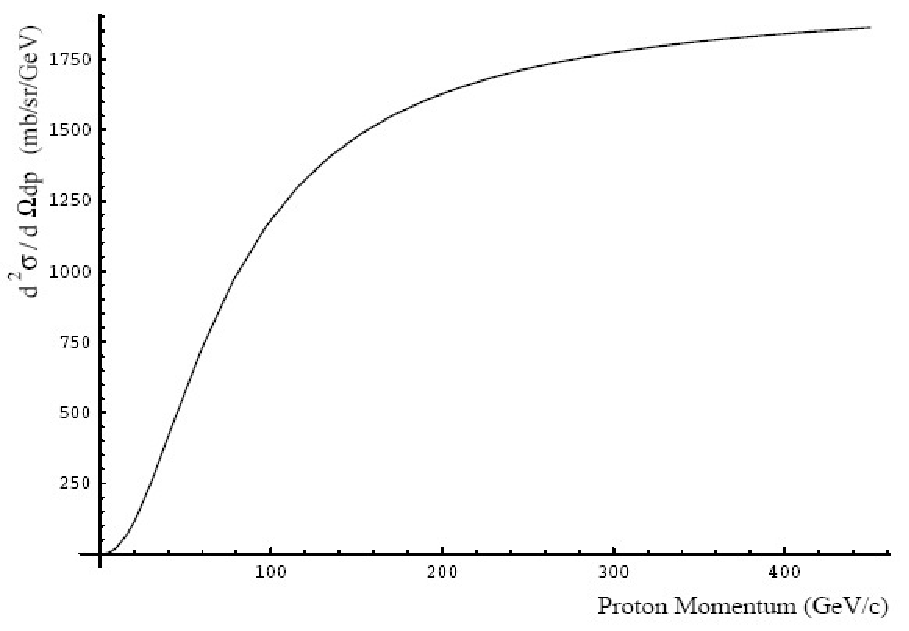,height=0.35\textheight}
\end{center}
\vskip -6mm
\caption{Differential cross-section for the production of $K_{\rm L}^{0}$ as a
function of the incident proton momentum \cite{belyaev}. The cross-section is
for a beryllium target in the forward direction, and with kaon momenta
integrated from 2 to 20 GeV. The estimate is based on the empirical Sanford-Wang
formula, using ${\rm d}\sigma (K_{\rm L}^{0}) = [{\rm d}\sigma (K^{+}) +
3{\rm d}\sigma (K^{-})]/4$.}
\label{fig:Kaon}
\end{figure}

By measuring the branching ratios of both $K \rightarrow \pi\nu\bar{\nu}$ decay
modes, the {\em unitarity triangle} of the CKM matrix can be completely
determined (see Fig.\,\ref{fig:Unitarity}), provided the matrix element $V_{cb}$
and the top quark mass are known \cite{buras}. Of particular interest is the
unitarity triangle parameter $\sin 2\beta$, which can also be determined from
the decay $B_{d} \rightarrow  \Psi{\rm K}_{s}$. Both determinations of this
parameter have to coincide if the Standard Model is valid \cite{belyaev}.

The decay $K_{\rm L}^{0} \rightarrow \pi^{0}\nu\bar{\nu}$ has not yet been 
observed. The KOTO experiment at J-PARC \cite{yamanaka}, the aim of which is to
study this decay mode, had its first physics run in May 2013. The current
branching ratio measurement of the charged decay mode, ${\cal B}(K^{+}
\rightarrow \pi^{+}\nu\bar{\nu}) = (17.3^{+11.5}_{-10.5})\times 10^{-11}$, is
based on the seven candidate events observed by the experiment E787/E949 at
Brookhaven \cite{artamonov}. This result is consistent with $(7.8\pm 0.80)\times
10^{-11}$, the value predicted by the SM \cite{brod}. The proposed  ORKA
experiment at Fermilab will use the stopped-kaon technique of its predecessor
E787/E949 to detect about 1000 $K^{+} \rightarrow \pi^{+}\nu\bar{\nu}$ decays,
and measure the corresponding branching ratio with a precision of 5\%
\cite{ritchie}. The NA62 experiment at CERN will rely on a complementary
decay-in-flight technique to detect about 100 $K^{+} \rightarrow \pi^{+}\nu
\bar{\nu}$ decays \cite{NA62}.

As shown in Fig.\,\ref{fig:Kaon}, the kaon yield rises rapidly as a function of
the incident proton momentum. From the figure one infers that the minimum energy
of the proton beam should be about 20 GeV, for otherwise the kaon yield would be
severely reduced. At the proposed KEK facility, a 70\,GeV proton synchrotron
could be installed, at a later stage, inside the Tristan ring in order to
increase the proton beam energy --- albeit at the cost of a considerably lower
beam power. For a given kaon yield, the required beam power would be lowest at
energies between 30 and 100 GeV \cite{belyaev}.

\newpage

\subsection[A Novel $g$\,-2 Experiment with Ultra-Slow Muons]{A Novel
{\boldmath{$g$}}\,{\bf-2} Experiment with Ultra-Slow Muons}
\vspace*{0.3cm}

~~~~A charged elementary fermion  has a {\em magnetic dipole moment}
{\boldmath{$\mu$}} $= g_{s}^{~}(q/2m)${\boldmath{$s$}} aligned with its spin
{\boldmath{$s$}}. The proportionality constant $g_{s}^{~}$ is the Land\'{e}
$g$-factor, $q$ is the charge of the particle and $m$ is its mass. Dirac's
theory of the electron predicts that $g_{s}^{~} = 2$. For the {\em electron}
(e), {\em muon} ($\mu$) and {\em tau lepton} ($\tau$), this prediction differs from the observed value by a small fraction of a percent. The difference is the
{\em anomalous magnetic moment}; the anomaly is defined by $a \equiv (g_{s}^{~}
- 2)/2 \sim 10^{-3}$. In the Standard Model, three distinct classes of Feynman
diagrams contribute to the value of the anomaly for each lepton species: (1) the
dominant QED terms that contain only leptons and photons; (2) terms that involve
hadrons; and (3) electroweak terms containing the Higgs, W and Z bosons. The
muon anomaly $a_{\mu}$ is about $(m_{\mu}/m_{e})^{2} \sim 43\,000$ times more
sensitive to the existence of yet unknown heavy particles than the electron
anomaly $a_{e}$. The value of $a_{\mu}$ ($a_{e}$) is sensitive to new physics at
the scale of a few hundred GeV (MeV) \cite{miller}.

\begin{figure}[h!]
\begin{center}
\epsfig{file=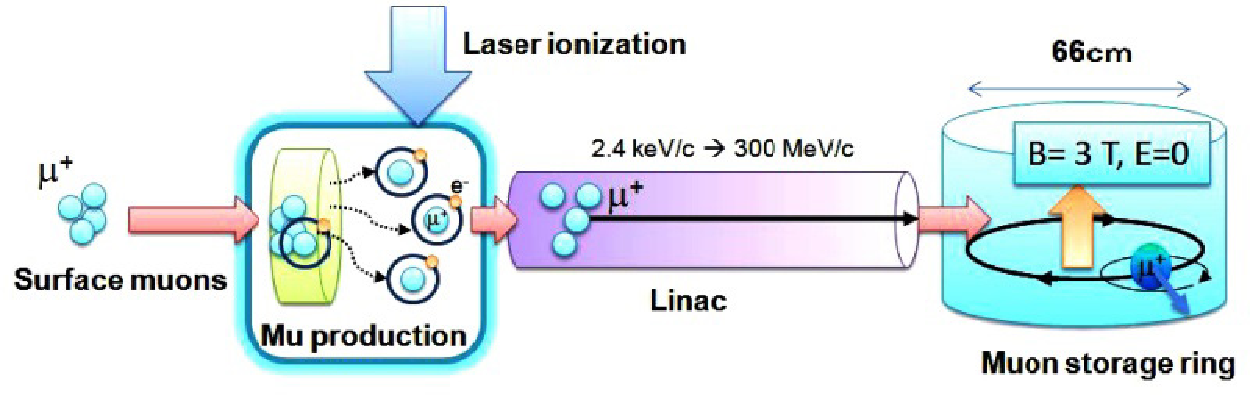,height=0.17\textheight}
\end{center}
\vskip -6mm
\caption{The conceptual design of a proposed experiment to measure the
{\em magnetic dipole moment} {\boldmath{$\mu$}}$_{\mu}$ $= g_{s}^{~}(q/
2m)${\boldmath{$s$}} and the {\em electric dipole moment} {\boldmath{$d$}}$_
{\mu}$ $= \eta(q\hbar /2m)${\boldmath{$s$}} of the muon \cite{aoki}.}
\label{fig:g-2}
\end{figure}

The current experimental uncertainty on $a_{\mu}$ is $\pm 0.54$ ppm. In a novel
$g$\,-2 experiment, the aim of which is to reduce this uncertainty to $\pm 0.1$
ppm (see Fig.\,\ref{fig:g-2}), 3\,GeV protons impinge on a graphite target and
produce pions that are stopped in the target. Some of the positive pions are
brought to rest near the surface of the target, where they decay into positive
muons with momenta $p_{\mu}^{~} = 30$\,MeV/c and 100\% spin polarization. The
muons are collected using a large-aperture solenoid and transported to a
silica-aerogel target in which they form muonium (electron--$\mu^{+}$) atoms. As
the atoms slowly diffuse from the target, they are ionized by a pulsed laser to
produce 50\% polarized muons with very low momenta.\footnote{A much higher level
of polarization can be obtained by using a magnetic field to align the particle
spins \cite{aoki}.} Those `ultra-slow' muons ($4\times 10^{4}$/pulse) are then
accelerated to $p_{\mu}^{~} = 300$\,MeV/c by two linacs, and injected into a
magnetic storage ring that contains a 3\,T solenoid with a diameter of 66\,cm.
After injection, the muons circulate orthogonal to the magnetic field {\bf B}.
An orbiting muon decays within 6.6 $\mu$s into a positron, a neutrino and an
antineutrino: $\mu^{+} \rightarrow {\rm e}^{+} + \nu_{e} + \bar{\nu}_{\mu}$.

The highest-energy positrons, preferentially emitted parallel to the muon spin
direction in the $\mu^{+}$ rest frame, are Lorentz-boosted to become the
highest-energy positrons in the lab frame. Hence, the angular distribution of
those positrons has its maximum in the direction of the muon spin
\cite{belusevic}. By measuring the energy and time distributions of positrons
one can determine the average spin direction. The time spectrum will show the
muon lifetime modulated by the spin precession frequency. The relative
precession of the spin with respect to the direction of the particle velocity
{\boldmath{$u$}} is given by {\boldmath{$\omega$}}$_{a}\,+${\boldmath{
$\omega$}}$_{\eta} \propto a_{\mu}{\bf B} - (\eta /2)$({\boldmath{$\beta$}}$
\times${\bf B}), where {\boldmath{$\omega$}}$_{a}$ and {\boldmath{$\omega$}}$
_{\eta}$ arise from $a_{\mu}$ and {\boldmath{$d$}}$_{\mu}$, respectively, and
{\boldmath{$\beta$}} $\equiv$ {\boldmath{$u$}}/$c$. Since the rotation axes due
to $a_{\mu}$ and  {\boldmath{$d$}}$_{\mu}$ are orthogonal, the corresponding
signals can be separated \cite{aoki}. In the case of {\boldmath{$\mu$}}$_{\mu}$,
the anomalous precession period is $2.2\,\mu$s, about 300 times the cyclotron
period. Assuming that muons are 100\% polarized, $1.5\times 10^{12}$ positrons
have to be detected for a measurement precision of 0.1 ppm \cite{aoki}.

\newpage

\section{An XFEL Based on the Proposed Superconducting Linac}
\vspace*{0.3cm}

~~~~To record the dynamics of atoms requires a probe with {\AA}ngstrom
($10^{-10}$ m) wavelength and  femtosecond temporal duration ($10^{-15}$ s).
Such probes have recently become available with the advent of {\em X-ray
free-electron lasers} (XFELs).\footnote{Optical lasers are capable of producing
pulses of femtosecond duration, but lack the required spatial resolution.
Due to their long pulse durations, X-rays from synchrotron light sources can be
used to image atomic structures only in static measurements.} The ultrashort
pulse duration of an XFEL matches the timescale of non-equilibrium microscopic
processes such as electron transfer in molecules, evolution of chemical
reactions, vibration dynamics in solid state systems, etc. Nanometer-scale
molecular imaging is made possible also by the high degree of {\em coherence}
of the XFEL radiation.
	
\begin{figure}[h]
\begin{center}
\epsfig{file=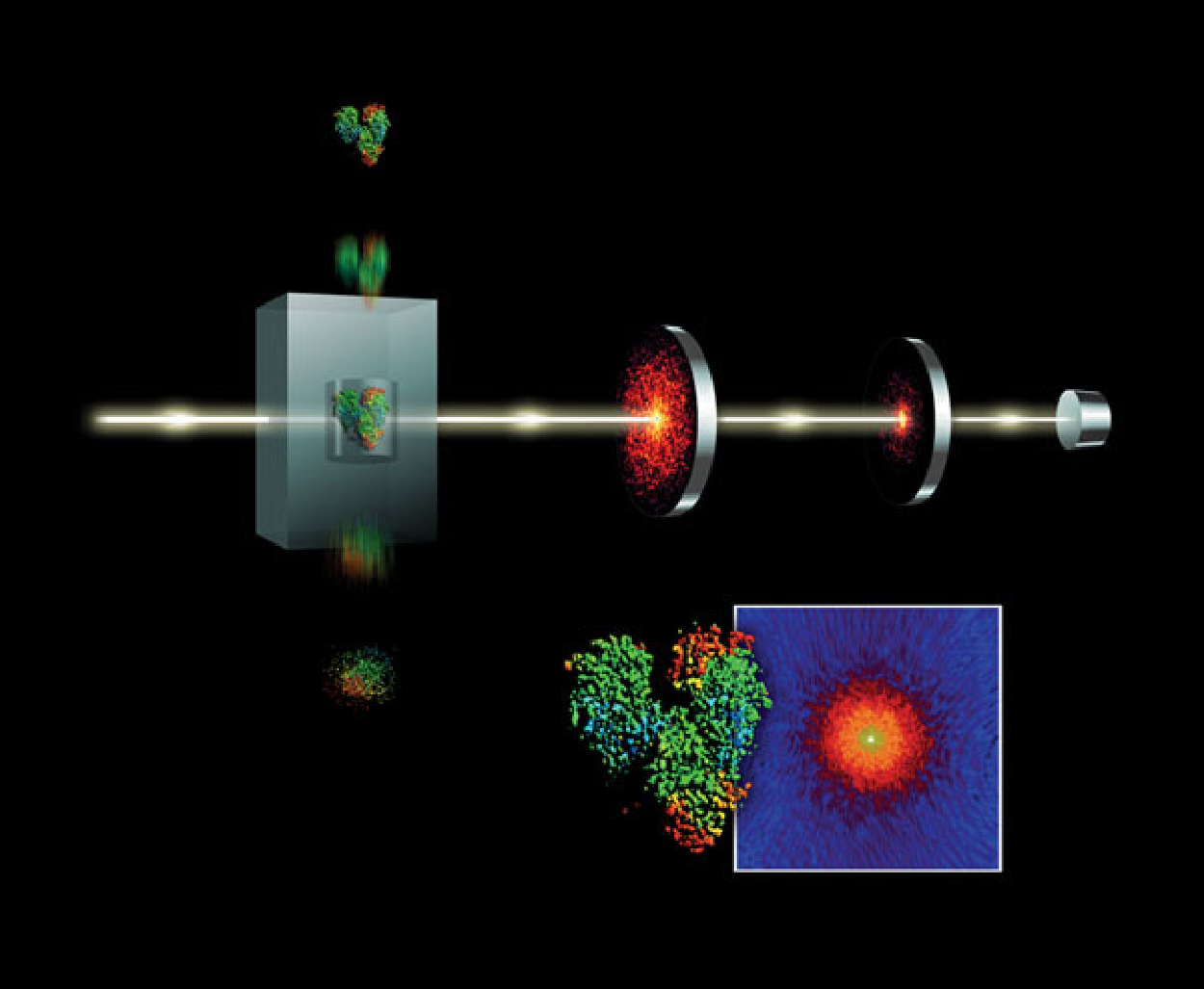,height=0.30\textheight}
\end{center}
\vskip -6mm
\caption{The LCLS records images of protein molecules as they fall through a
target chamber. The drawing shows how scattering of X-ray pulses with
femtosecond duration  records locations of individual atoms before the pulse
energy tears the protein apart; credit: LCLS and SLAC.}
\label{fig:Imaging}
\end{figure}

The peak {\em spectral brightness} of the two presently most powerful XFEL
facilities --- LCLS at SLAC (United States) and SACLA at SPring-8 (Japan) --- is
billion times higher than that of any synchrotron radiation source. Owing to the
high intensity of XFEL radiation, laser-irradiated atoms, molecules and atomic
clusters can be excited into previously unknown states. Although high-intensity
pulses may also destroy molecular structures, they can still be used to produce
high-resolution X-ray diffraction patterns (see Fig.\,\ref{fig:Imaging}), from
which real-space images of the atomic positions in molecules can be
reconstructed. In a typical `pump-probe' experiment, the evolution of a chemical
(or biochemical) reaction, initiated by an optical or IR laser pulse, is
observed by a time-delayed X-ray pulse. By varying the delay, such stroboscopic
measurements result in femtosecond `movies' of the evolving system.

\vspace*{0.3cm}
\subsection{A Simplified Description of X-Ray Free-Electron Lasers}
\vspace*{0.3cm}

~~~~Despite its name, the FEL is more closely related to vacuum tube devices
than lasers. Whereas in a conventional laser light amplification is created by
the stimulated emission of electrons bound to atoms, the amplification medium
of the FEL are `free' (unbound) electrons. Free-electron lasing is achieved by a
single-pass, high-gain FEL amplifier operating in the {\em self-amplified
spontaneous emission} (SASE) mode.

An FEL consists of an electron linear accelerator and an {\em undulator}, a long
periodic array of magnets with period $\lambda_{u}$. The undulator generates a
sinusoidal transverse magnetic field described by ${\rm B} = {\rm B}_{0}\sin
(2\pi z/\lambda_{u})$. In the rest frame of an electron, the magnetic field of
the undulator becomes a combination of a transverse magnetic field and a
transverse electric field, travelling together at almost the speed of light. The
electron, therefore, `sees' the undulator as an electromagnetic wave with
wavelength given by the undulator period corrected for the relativistic Lorentz
contraction: $\lambda^{*} = \lambda_{u}/\gamma$, where the Lorentz factor
$\gamma = {\rm E}_{e}/m_{e}c^{2}$ is defined as the relativistic energy of the
electron in units of its rest energy $m_{e}c^{2}$. This wave causes the electron
to oscillate as a classical radiating dipole and emit electromagnetic waves with
wavelength $\lambda^{*}$.\footnote{In the laboratory frame, electrons emit
radiation in the forward direction within a narrow cone of opening angle
$1/\gamma$. The cone is centred around the instantaneous tangent to the electron
trajectory. The direction of the tangent varies along the sinusoidal orbit in
an undulator, the maximum angle with respect to the $z$-axis being $\theta_
{\rm max} \sim K/\gamma$, where $K \propto {\rm B}_{0}\lambda_{u}$. If
$\theta_{\rm max} \leq 1/\gamma$, the radiation field contributions from various
sections of the trajectory overlap in space and interfere with each other.
Consequently, the radiation spectrum at $\theta = 0$ is nearly monochromatic,
and the {\em angular width} of the {\em first harmonic} is $\sigma_{\theta}^{~}
\sim 1/\gamma\sqrt{N_{u}}$, where $N_{u}$ is the number of undulator
periods \cite{schmuser}.} In the laboratory frame, the wavelength of the
radiation is Doppler-shifted: $\lambda =\lambda^{*}\gamma (1 - \beta\cos\theta )
\approx (\lambda_{u}/2\gamma^{2})(1 + \gamma^{2}\theta^{2})$, where $\theta$ is
the angle with respect to the forward direction and $\gamma^{2} =(1 - \beta^{2})
^{-1}$. Taking into account the reduced longitudinal electron velocity caused by
the transverse motion (responsible for the second term in Eq. (8)), the
{\em wavelength} of the {\em first harmonic} of the observed radiation is given
by \cite{schmuser, huang}
\begin{equation}
\lambda = \frac{\raisebox{-.4ex}{$\lambda_{u}$}}{2\gamma^{2}}\!\left (
1 + \frac{\raisebox{-.4ex}{$K^{2}$}}{2} + \gamma^{2}\theta^{2}\right )
\end{equation}
where the dimensionless quantity
\begin{equation}
K = \frac{\raisebox{-.4ex}{$e{\rm B}_{0}\lambda_{u}$}}{2\pi m_{e}c} =0.934\,
{\rm B}_{0}[\mbox{Tesla}]\,\lambda_{u}[\mbox{cm}]
\end{equation}
is the undulator {\em deflection parameter}. Typically, $\lambda_{u} \approx 3$
cm and $\gamma \approx 10^{4}$; hence, $\lambda \approx 0.1$ nm.

\begin{figure}[t]
\vspace{-3.3mm}
\begin{center}
\epsfig{file=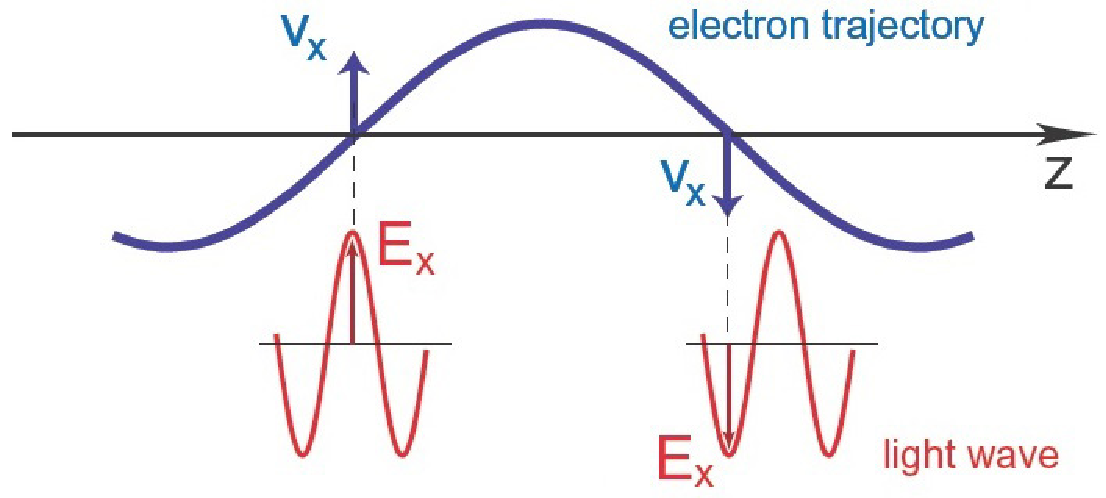,height=0.20\textheight}
\end{center}
\vskip -5mm
\caption{Condition for energy transfer from an electron to the radiation field
in an undulator. The electromagnetic wave, which propagates with the speed of
light and hence always moves ahead of an electron, has to `advance' by half an
optical wavelength in a half-period of the electron trajectory \cite{schmuser}.}
\label{fig:XFEL}
\end{figure}

The optical amplification in an XFEL is caused by a sustained energy transfer
from the electrons to the co-moving radiation field. This energy transfer
can only take place if the transverse velocity component of an electron and the
electric vector of the electromagnetic wave point in the same direction. For
this condition to be satisfied, the electromagnetic wave has to `advance' by the
right amount (see Fig.\,\ref{fig:XFEL}), and this is only possible for a certain
wavelength $\lambda_{\ell}$. A simple calculation shows that $\lambda_{\ell} =
\lambda (\theta = 0)$, where $\lambda (\theta )$ is given by Eq. (8).

\begin{figure}[t]
\vspace{-3.3mm}
\begin{center}
\epsfig{file=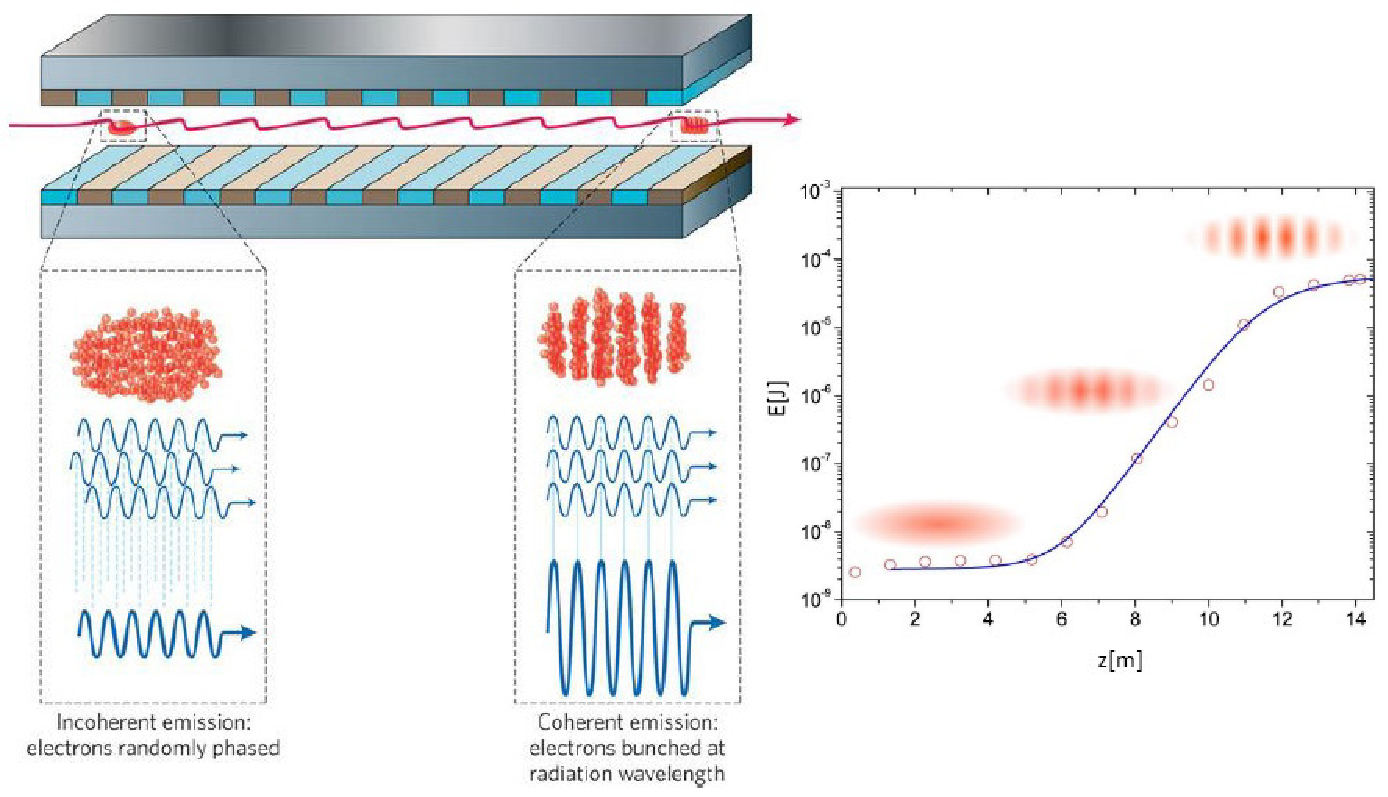,height=0.36\textheight}
\end{center}
\vskip -6mm
\caption{Electrons entering an undulator have random phases and thus initially
emit mostly incoherent radiation at the wavelength $\lambda$. Since the
electrons interact collectively with the radiation they emit, small coherent
fluctuations in the radiation field grow and simultaneously begin to bunch the
electrons. This collective process continues until the electrons are strongly
bunched \cite{McNeil} (left figure). The progress of microbunching and the
exponential growth of the FEL pulse energy as a function of the distance $z$
traversed along the undulator  are shown in the figure on the right. Open
circles represent a measurement \cite{FLASH}, and the solid curve is a
theoretical prediction \cite{schmuser}.}
\label{fig:SASE}
\end{figure}

Initially, an electron bunch in the undulator is much longer than the radiation
wavelength, and the electrons are distributed uniformly throughout the bunch.
Depending on the relative phase between radiation and electron oscillation, some
electrons will gain energy from the radiation field while other electrons will
lose energy to the field. As faster electrons catch up with the slower ones, a
periodic density modulation on the scale of $\lambda_{\ell}$ (the so-called
{\em microbunching}) begins to develop in the undulator. Since these
microbunches are close to the positions where maximum energy transfer to the radiation field takes place, the microbunched electron beam emits coherent
radiation at the expense of the beam energy. The increase in the radiation field
enhances the microbunching even further, which leads to an {\em exponential
growth} of the FEL pulse energy as a function of the distance $z$ traversed
along the undulator: ${\cal P}(z) \propto {\rm e}^{z/L_{\rm g}}$, where
${\cal P}$ denotes power and $L_{\rm g} \approx 50\lambda_{u}$ is the {\em power
gain length} (see Fig.\,\ref{fig:SASE}). The exponential power growth lasts
until both the radiation intensity and the electron beam microbunching reach a
{\em saturation level} (at $L_{\rm sat} \approx 20L_{\rm g}$). This occurs when
the beam loses so much energy that the resonant condition is no more satisfied
\cite{schmuser}.

The total electric field of the undulator radiation is the sum of the fields
from all the electrons: ${\rm E}_{\rm tot} = \sum_{n}{\rm E}_{n}{\rm e}^
{i\phi_{n}}$, where E$_{n}$ is the field due to a single electron and $\phi_{n}$
is the phase of the electric field from the $n^{\rm th}$ electron. The power
emitted by the electrons is given by
\begin{equation}
{\cal P} \propto \left |\sum_{n}{\rm E}_{n}{\rm e}^{i\phi_{n}}\right |^{2} =
\sum_{n}{\rm E}_{n}^{2} \,+\,\left |\sum_{n}\sum_{m}{\rm E}_{n}{\rm E}_{m}\,
{\rm e}^{i(\phi_{n}\,-\,\phi_{m})}\right |_{n\,\neq\,m}
\end{equation}
Since the $N_{e} \sim 10^{7}$ electrons within a microbunch oscillate in phase,
the individual fields add coherently. In this case $\phi_{n}$ is a constant and
$|{\rm E}_{\rm tot}| \approx n_{b}^{~}N_{e}{\rm E}_{0}$; here $n_{b}^{~} \sim
10^{2}$ is the number of microbunches and ${\rm E}_{n}$ was set equal to
${\rm E}_{0}$. The dominant contribution to the total {\em coherently emitted}
power comes from the second term in Eq. (10): ${\cal P}_{\rm coh} \approx n_{b}
^{~}(N_{e}{\rm E}_{0})^{2}$. If the radiation is produced by particles
oscillating at random phases ({\em incoherent emission}), the second sum in Eq.
(10) tends to interfere destructively (summing the fields is equivalent to a
random walk in the complex plane). In this case the dominant
first term gives ${\cal P}_{\rm incoh} \approx n_{b}^{~}N_{e}{\rm E}_{0}^{2}
\ll {\cal P}_{\rm coh}$ \cite{McNeil}.

The emission of radiation in an undulator does not occur at one wavelength, but
in a {\em wavelength band} of width $\Delta\lambda$ around the central value
given by Eq. (8). Each electron propagating through the undulator emits a wave
train consisting of a number of wavelengths equal to the number of undulator
periods, $N_{u}$. The time duration $\Delta t$ of this pulse is the {\em pulse
length} $L_{p} \equiv N_{u}\lambda$ divided by the speed of light: $\Delta t =
N_{u}\lambda /c$. A pulse of duration $\Delta t$ has a {\em frequency bandwidth}
$\Delta\nu \sim 1/\Delta t$. Hence, $\Delta\nu \sim c/N_{u}\lambda = \nu /
N_{u}$, because $\lambda = c/\nu$. Thus,
\begin{equation}
\frac{\raisebox{-.4ex}{$\Delta\nu$}}{\raisebox{.4ex}{$\nu$}} = \frac{\raisebox
{-.4ex}{$\Delta\lambda$}}{\lambda} \sim \frac{\raisebox{-.4ex}{1}}{N_{u}}
\approx 10^{-3}
\end{equation}
The wave train is not monochromatic due to its finite length. For typical
values $N_{u} \approx 10^{3}$ and $\lambda \approx 0.1$ nm, one obtains
$\Delta t \approx 0.33$ fs. Since the electrons are distributed throughout a
bunch, the {\em pulse duration} is increased to $\tau_{p}\sim (\sigma_{z}/L_{p})
\Delta t \approx 80$ fs, where $\sigma_{z} \approx 25~\mu$m is the {\em bunch
length} and $L_{p} \approx 0.1~\mu$m. An 80\,fs pulse, therefore, consists of
many micropulses of 0.33\,fs duration.

The `shot noise' in an electron beam, the origin of which is the random emission
of the electrons from a photocathode (see Fig.\,\ref{fig:Euro}), causes random
fluctuations of the beam density. The radiation produced by such a beam has
amplitudes and phases that are random in both space and time. For this reason,
SASE X-ray FELs lack {\em longitudinal} (or {\em temporal}) {\em coherence},
characterized by the {\em coherence length} $L_{\rm coh}\equiv\lambda^{2}/\Delta
\lambda \approx 0.1~\mu$m. This quantity is defined as the distance of
propagation over which radiation with spectral width $\Delta\lambda$ becomes
180$^{\circ}$ out of phase.\footnote{For a wavelength $\lambda$ propagating
through $n$ cycles, $L_{\rm coh} = n\lambda$; for a wavelength $\lambda + \Delta
\lambda$ propagating through $(n - 1/2)$ cycles, $L_{\rm coh} = (n - 1/2)
(\lambda + \Delta\lambda )$. Hence $L_{\rm coh} \approx \lambda^{2}/2\Delta
\lambda$, although $L_{\rm coh} \equiv \lambda^{2}/\Delta\lambda$ is also often
used.} The {\em coherence time}, defined by $t_{\rm coh}\equiv L_{\rm coh}/c
\sim 1/\Delta\nu$, is much shorter than the {\em pulse duration}:
$t_{\rm coh} \approx 0.3$ fs.

In order to increase the coherence length in the hard X-ray regime (photons with
0.1 nm wavelength), a `{\em self-seeding}' method was tested at LCLS
\cite{amann}. FEL pulses, generated in the first modular section of the LCLS
undulator, are spectrally `purified' by a crystal filter (a diamond
monochromator). Since a typical monochromator delays X-rays, the electron
bunches exiting the first modular section are appropriately delayed after being
diverted around the crystal by a compact magnetic chicane (see Fig.\,1 in
\cite{amann}). The crystal selects a very narrow part of the spectrum, which is
further amplified in the second undulator section where the FEL radiation
reaches saturation. At LCLS, `self-seeding' generated X-ray pulses with $\Delta
\nu =$ 0.4--0.5 eV at $\nu =$ 8--9 keV, which represents a factor of 40--50
{\em bandwidth reduction} with respect to SASE \cite{amann}.
	
\vspace*{0.3cm}
\subsection{The European XFEL as a Prototype of the Proposed X-Ray FEL}
\vspace*{0.3cm}

\begin{figure}[t]
\vspace{-3.3mm}
\begin{center}
\epsfig{file=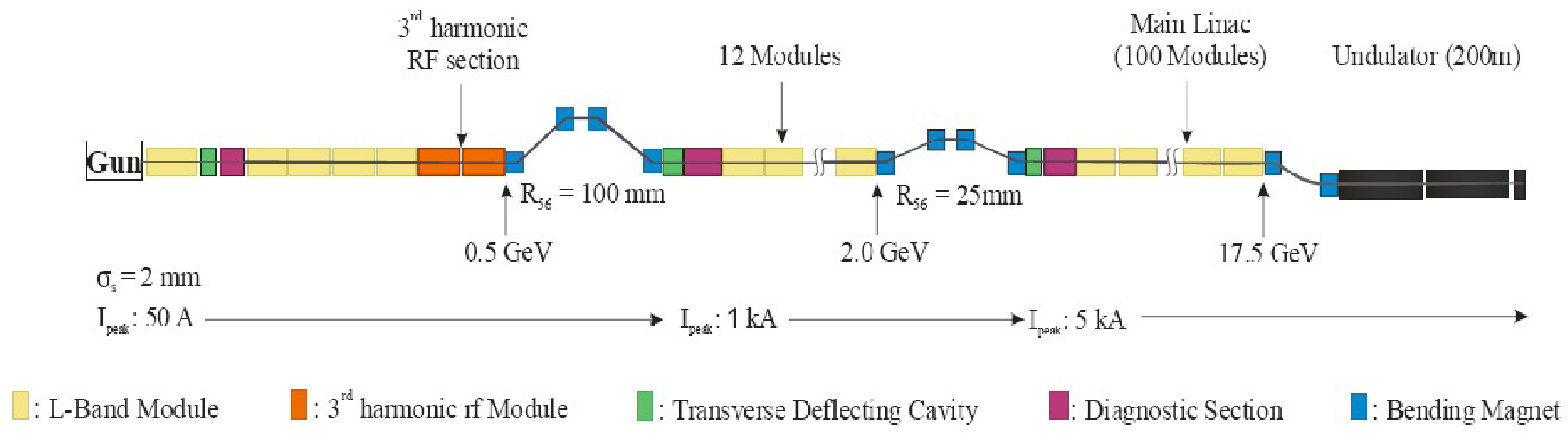,height=0.18\textheight}
\end{center}
\vskip -6mm
\caption{Schematic layout of the European XFEL \cite{TDR}. Electron bunches,
each with a charge of 1\,nC, are extracted from a photocathode by short
ultraviolet laser pulses and then focused and accelerated inside a
radio-frequency cavity (`RF gun') to an energy of 120 MeV. In order to produce
5\,kA peak currents necessary for lasing, the bunches are further accelerated
and longitudinally compressed down to 25~$\mu$m using two magnetic chicanes (at
0.5 and 2.0 GeV). After traversing the main linac, where their energy is
increased to 17.5\,GeV, the bunches enter FEL undulators.}
\label{fig:Euro}
\end{figure}

~~~~The European XFEL, currently under construction at DESY (Germany), is a
free-electron laser (FEL) based on self-amplified spontaneous emission (SASE)
in the X-ray regime. The FEL consists of a 17.5\,GeV superconducting electron
linear accelerator and a set of undulators (see Fig.\,\ref{fig:Euro}) that can
generate both SASE FEL X-rays and incoherent radiation. A superconducting linac
may accelerate 10 `bunch trains' per second, each train consisting of up to 2700
electron bunches. This results in 27\,000 ultrashort X-ray flashes per second
--- many more than at any other existing XFEL facility. The higher the number of
electron bunches, the more scientific instruments can be operated
simultaneously. The European XFEL facility will generate ultra-short pulses
($\leq 100$ fs) of spatially and temporally coherent X-rays with wavelengths in
the range $\sim$\,0.1--5 nm.

The {\em spectral brightness} (or {\em brilliance}), $\frak{B}$, of a radiation
field is defined as the number of photons per unit phase-space volume per unit
fractional bandwidth per unit time. In practical units, $\frak{B}$ can be
expressed as
\begin{equation}
\mbox{Spectral brightness} \equiv \frac{\raisebox{-.4ex}{Number of photons}}
{({\rm second})({\rm mm}^{2})({\rm mrad}^{2})(0.1\%~{\rm BW})}
\end{equation}
where BW denotes {\em spectral bandwidth}. The quantity $\frak{B}$ determines how
much monochromatic radiation can be focused onto a tiny spot on the target. The
{\em peak spectral brightness} of the FEL is the brightness measured during the
very short duration of an FEL pulse. The peak brilliance of the European XFEL is
expected to be $5\times 10^{33}$ photons/second/mm$^{2}$/mrad$^{2}$/0.1\%~BW.

\begin{figure}[h!]
\begin{center}
\epsfig{file=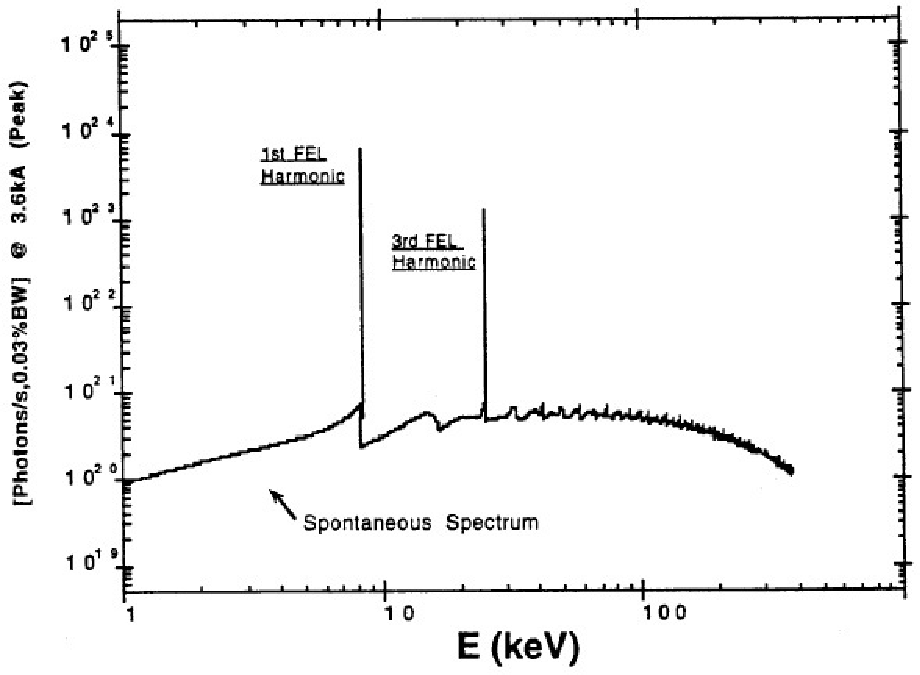,height=0.3\textheight}
\end{center}
\vskip -6mm
\caption{Computed spectral flux of spontaneous undulator and FEL radiation at
LCLS. Plotted is the number of photons per second and 0.3\% bandwidth as a
function of photon energy \cite{arthur}.}
\label{fig:Spectrum}
\end{figure}

As mentioned earlier, the coherent superposition of the radiation fields from
{\em all} microbunches is responsible for the nearly monochromatic spectrum and
small divergence of the radiation emitted in the forward direction (see footnote
5 and \cite{schmuser}). Recall also that `self-seeding' can substantially
improve longitudinal (temporal) coherence of SASE XFEL radiation (see Section
4.1). Thus, the radiation from an X-ray FEL has a narrow bandwidth, is
transversely and longitudinally coherent, and is fully polarized. The coherently
emitted XFEL spectral lines appear in addition to the spontaneously
emitted undulator spectrum that extends into the MeV energy region (see
Fig.\,\ref{fig:Spectrum}).

\begin{figure}[t]
\vspace{-3.3mm}
\begin{center}
\epsfig{file=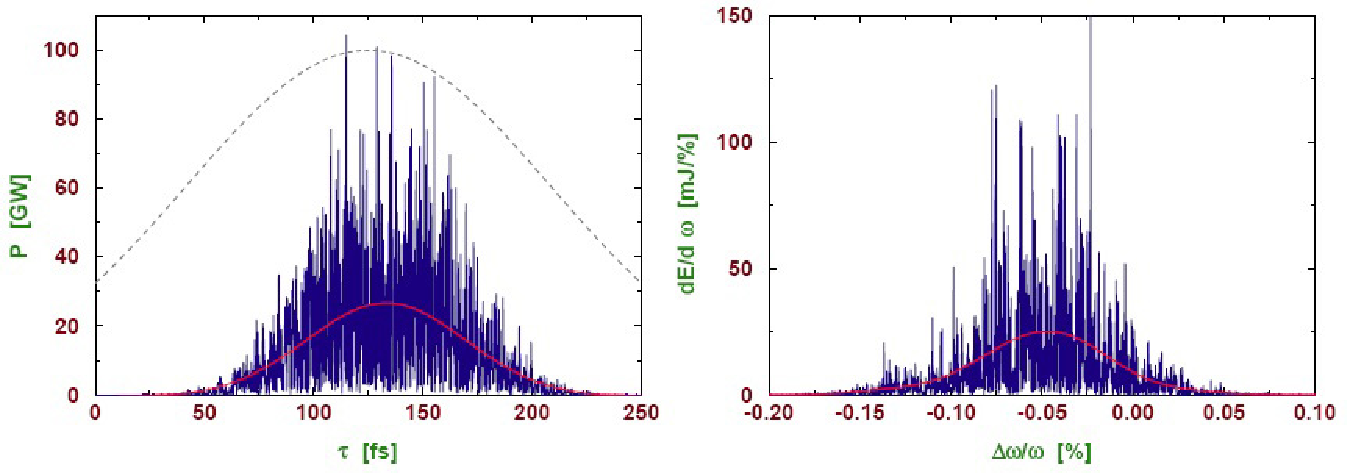,height=0.23\textheight}
\end{center}
\vskip -6mm
\caption{Typical temporal (left) and spectral (right) structure of the radiation
pulse from a SASE XFEL at a wavelength of 0.1 nm \cite{Tesla}. The red lines
correspond to average values. The dashed line represents the axial density
profile of the electron bunch. The line width is inversely proportional to the
coherence time.}
\label{fig:Spikes}
\end{figure}

The micropulses that form an FEL pulse give rise to `spikes' shown in
Fig.\,\ref{fig:Spikes}. The amplitudes of the micropulses vary greatly as a
consequence of the amplified stochastic variations in the electron density. Within a micropulse, the radiation is both transversely and longitudinally
coherent. The duration of a micropulse is roughly $t_{\rm coh}$, the
{\em coherence time}. In the SASE\,1 and SASE\,2 undulators at the European
XFEL, $t_{\rm coh} =$ 0.2--0.38 fs \cite{TDR}. The number of `spikes' in a pulse
is given by the ratio of the {\em bunch length} to the {\em coherence length}:
$\sigma_{z}/L_{\rm coh} = (25~\mu{\rm m})/(0.1~\mu{\rm m}) \approx 250$.

The spectrum of undulator radiation is sharply peaked around {\em odd
harmonics}\,\footnote{The occurrence of {\em higher harmonics} is explained in
\cite{schmuser}. In the forward region ($\theta = 0$) of a planar undulator,
only the odd higher harmonics are observed, while the off-axis radiation
contains also the even harmonics. For the European XFEL, simulations predict that the relative contribution to the total radiation power of the 3$^{rd}$ and
the 5$^{th}$ harmonic is about 1\% and 0.03\%, respectively \cite{TDR}.} (see
Fig.\,\ref{fig:Spectrum}). The photon energy that corresponds to the
$n^{\rm th}$ harmonic is given by
\begin{equation}
{\rm E}_{n}\,[{\rm keV}] \,=\, 0.9496\,\frac{\raisebox{-.4ex}{$n{\rm E}_{e}^{2}
\,[{\rm GeV}]$}}{\lambda_{u}\,[{\rm cm}](1 + K^{2}/2 + \gamma^{2}\theta^{2})}
\end{equation}
where E$_{e}$ is the electron beam energy and $\theta$ is the radiation
detection angle (with respect to the forward direction). For $\theta = 0$ and
the SASE\,2 undulator parameters $\lambda_{u} = 4.8$ cm and $K = 6.1$, for
example, Eq. (13) yields E$_{1} = 12.2$ keV \cite{TDR}.

At the exit of the SASE\,1 undulator, the photon beam divergence is
$\sigma_{\theta}^{~}\sim 1/\gamma\sqrt{N_{u}}\approx 1\,\mu$rad (see footnote 5)
and the beam size is $70\,\mu{\rm m}\times 70\,\mu{\rm m}$, the diameter of a
fine needle. This beam can be focused to an area of $0.1\,\mu{\rm m}\times 0.1\,
\mu{\rm m}$ (the size of a virus) at an experimental station located a couple of
hundred meters from the undulator exit \cite{agapov}. Through variable focusing,
the flux density of an XFEL beam can therefore be tuned by a factor of about one
million. The SASE\,1 undulator will deliver $10^{12}$ photons in an ultra-short
pulse of 100\,fs duration (the timescale of molecular vibrations), yielding a
peak power of about 20 GW at a photon energy ${\rm E} \approx 12$ KeV ($\lambda =
hc/{\rm E} \approx 0.1$\,nm).

As already mentioned, nanometre-scale molecular imaging is made possible by the
high degree of coherence of the XFEL radiation. The coherence quality of a light
source is best described by the {\em degeneracy parameter} $\frak{D}$, defined as
the number of photons per coherent phase-space volume $(\lambda /2)^{2}$ per
coherence time $c^{-1}(\lambda^{2}/\Delta\lambda )$. Since the unit fractional
bandwidth is $\Delta\lambda /\lambda$,
\begin{equation}
\frak{D} \equiv \frak{B}\!\left (\frac{\raisebox{-.4ex}{$\lambda$}}{2}\right )
^{2}\!\left (\frac{\raisebox{-.4ex}{$\lambda^{2}/\Delta\lambda$}}{c}\right )
\!\left (\frac{\raisebox{-.4ex}{$\Delta\lambda$}}{\lambda}\right )\! =
\frac{\raisebox{-.4ex}{$\frak{B}\lambda^{3}$}}{4c} \approx 8.3\times 10^{-25}\,
\frak{B}\lambda^{3}
\end{equation}
Here $\frak{B}$ is the {\em brilliance} and $\lambda$\,[0.1\,nm] the wavelength
of the source. Recall that all photons in a single micropulse are
completely coherent. Since each pulse contains $\sim 10^{12}$ photons and a few
hundred micropulses, there are $10^{9}$ indistinguishable (`degenerate') photons
in the coherence volume. In comparison, $\frak{D} \approx 0.03$ at a synchrotron
source with $\lambda = 0.1$\,nm. Because of the large transverse coherence area
of $70\times 70\,\mu{\rm m}^{2}$ and the large number of coherent photons per
pulse, an interference (`speckle') pattern can be recorded with a single XFEL
pulse \cite{pellegrini, barty}.

\newpage

\section{Summary and Acknowledgements}
\vspace*{0.3cm}

~~~~The main `bottleneck' limiting the beam power in circular machines is caused
by space charge effects that produce beam instabilities. Such a `bottleneck'
exists at the J-PARC proton synchrotron complex, and is also intrinsic to the
`proton drivers' envisaged at CERN and Fermilab. In order to maximally increase
the beam power of a `proton driver', it is proposed to build a facility
consisting solely of a low-energy injector linac (PI) and a high-energy
{\em pulsed} superconducting linac (SCL). The 2.5\,GeV PI would serve both as an
injector to the SCL and a source of proton beams that could be used to copiously
produce neutrons and muons. Protons accelerated by the SCL to 20 GeV would be
transferred through the KEK Tristan ring in order to create neutrino, kaon and
muon beams for fixed-target experiments. At a later stage, a 70\,GeV proton
synchrotron could be installed inside the Tristan ring. The proposed facility
would be constructed using the existing KEK accelerator infrastructure.

High-power proton linear accelerators have a wide range of applications
including spallation neutron sources, nuclear waste transmutation, production of
radioisotopes for medical use, etc. A number of laboratories worldwide have
expressed interest in building `proton drivers' that would primarily deliver
high-intensity neutrino, kaon and muon beams.

Experiments with high-intensity neutrino beams are designed primarily to explore
the mass spectrum of the neutrinos and their properties under the CP symmetry.
The proposed T2HK (Tokai-to-Hyper-Kamiokande) project, for example, is a natural
extension of the successful T2K (Tokai-to-Super-Kamiokande) long-baseline
neutrino oscillation experiment. Hyper-Kamiokande (HK), a water Cherenkov
detector with a fiducial mass of 0.54 million metric tons, would serve as a
`far' detector for neutrino beams produced at J-PARC, situated 295 km away from
Kamioka.

{\em In case} HK {\em is never built, the proposed} `{\em proton driver}'
{\em and the existing} Super-Kamiokande {\em detector would produce, within a
given period of time, roughly the same number of $\nu_{e}^{~}$ and $\bar{\nu}_
{e}^{~}$} `{\em appearance events}' {\em as the} T2HK {\em experiment}.
Alternatively, a 100\,kiloton water Cherenkov detector could be built at
Okinoshima, located along the T2K beamline at a distance of about 650 km from
KEK. Using the proposed `proton driver', the detector at Okinoshima and
Super-Kamiokande, one could determine the neutrino mass hierarchy as well as
measure the CP-violating phase in the neutrino mixing matrix. To produce
neutrino beams, a DC magnetic horn would be employed.

Some of the most important discoveries in particle physics emerged from
high-precision studies of K mesons (`kaons'), in particular {\em neutral
kaons}. A deeper insight into CP violation is expected to be gained from
measurements of ultra-rare kaon decays such as $K_{\rm L}^{0} \rightarrow
\pi^{0}\nu\bar{\nu}$ and $K_{\rm L}^{+} \rightarrow \pi^{+}\nu\bar{\nu}$. These
decays provide important information on higher-order effects in electroweak
interactions, and therefore can serve as a probe of new phenomena not predicted
by the Standard Model.

A unique feature of the proposed facility is the use of superconducting ILC-type
cavities to accelerate {\em both} protons and electrons, which considerably
increases its physics potential. Polarized electrons and positrons can be
employed to study the structure of composite particles and the dynamics of
strong interactions, as well as to search for new physics beyond the Standard
Model.

An SCL-based X-ray free-electron laser (XFEL) and a synchrotron light source
for applications in materials science and medicine are also envisaged. The
ultrashort pulse duration of an XFEL matches the timescale of non-equilibrium
microscopic processes, allowing the dynamics of atoms and molecules to be
recorded in the form of femtosecond `movies'.

\vspace*{0.5cm}
\begin{center}
{\large\bf Acknowledgements}
\end{center}
\vspace*{0.3cm}

For many useful discussions regarding various aspects of this proposal, I am
very grateful to K. Fujii, K. Hagiwara, T. Higo, E. Kako, Y. Kamiya, K. Oide,
K. Takayama and K. Tokoshuku. I wish to express my special gratitude to 
Kaoru Yokoya for the invaluable help and encouragement I have received from him
over the years.

\end{document}